\documentclass[prd,aps,twocolumn,nofootinbib]{revtex4-2}

\usepackage{amsfonts,amsmath,amssymb}
\usepackage[colorlinks=true, citecolor=blue, linkcolor=blue, urlcolor=blue]{hyperref}
\usepackage{graphicx,multirow}
\usepackage{color}

\newcommand{\DS}[1]{/\!\!\!#1}

\allowdisplaybreaks

\begin{document}

\title{$\eta_c$ leading-twist distribution amplitude and the $B_c \to \eta_c\ell\bar\nu_\ell$ semileptonic decays using QCD Sum Rules}

\author{Long Zeng}
\email{zlong@cqu.edu.cn}
\author{Xing-Gang Wu}
\email{wuxg@cqu.edu.cn}
\author{Dan-Dan Hu}
\email{hudd@stu.cqu.edu.cn}
\author{Yu-Jie Zhang}
\email{zhangyj@stu.cqu.edu.cn}
\address{Department of Physics, Chongqing Key Laboratory for Strongly Coupled Physics, Chongqing University, Chongqing 401331, P.R. China}

\author{Hai-Bing Fu}
\email{fuhb@gzmu.edu.cn}
\author{Tao Zhong}
\email{zhongtao1219@sina.com}
\address{Department of Physics, Guizhou Minzu University, Guiyang 550025, P.R. China}
\address{Institute of High Energy Physics, Chinese Academy of Sciences, Beijing 100049, P.R.China}

\date{\today}

\begin{abstract}

In this paper, we investigate the semileptonic decays $B_c \to \eta_c\ell\bar\nu_\ell$ using the quantum chromodynamics(QCD) sum rules within the framework of Standard Model (SM). We further explore the potential to probe signatures of new Physics (NP) beyond the SM through these decays. First, we derive the $\xi$-moments $\langle\xi_{2;\eta_c}^{n}\rangle$ of the $\eta_c$-meson leading-twist distribution amplitude $\phi_{2;\eta_c}$ using the QCD sum rules within the background field theory. Considering contributions from the vacuum condensates up to dimension-six, the first two nonzero $\xi$-moments at the scale of $4$ GeV are found to be $\langle\xi_{2;\eta_c}^{2}\rangle = 0.103^{+0.009}_{-0.009}$ and $\langle\xi_{2;\eta_c}^{4}\rangle = 0.031^{+0.003}_{-0.003}$. Using these moments, we then fix the Gegenbauer expansion series of $\phi_{2;\eta_c}$ and apply it to compute the $B_c \to \eta_c$ transition form factors (TFFs) using QCD light cone sum rules. Second, we extrapolate those TFFs to physically allowable $q^2$-range via a simplified series expansion, and we obtain $R_{\eta_c}|_{\rm SM} = 0.308^{+0.084}_{-0.062}$. Furthermore, we explore the potential impacts of various NP scenarios on $R_{\eta_c}$. Specifically, we compute the forward-backward asymmetry $\mathcal{A}_{\rm FB}({q^2})$, the convexity parameter $\mathcal{C}_F^\tau ({q^2})$, and the longitudinal and transverse polarizations $\mathcal{P}_L ({q^2})$ and $\mathcal{P}_T ({q^2})$ for $B_c \to \eta_c$ transitions within both the SM and two types of NP scenarios. Our results contribute to a deeper understanding of $B_c$-meson semileptonic decays and provide insights into the search for the NP beyond the SM.

\end{abstract}

\maketitle

\section{Introduction}

Semileptonic $b$-hadron decays are potent instruments for testing the Standard Model (SM) and searching for signatures of new physics (NP) beyond the SM. These decays serve as an ideal platform for testing the universality of charged-lepton couplings in electro-weak interactions. This is because their theoretical predictions in the SM rely on relatively straightforward tree-level calculations. In recent years, tensions between experimental measurements and Standard Model predictions have emerged in the flavor sector, particularly in the $b\to c\ell\nu$ and $b\to s\ell\nu$ flavor-changing processes. These so-called flavor anomalies could serve as compelling evidence for NP if confirmed. Several measurements of the branching ratios $R_{D^{(*)}}$ and $R_{J/\psi}$ have been done by the BABAR~\cite{BaBar:2013mob}, the Belle~\cite{Belle:2015qfa, Belle:2016ure, Belle:2016dyj}, and the LHCb~\cite{LHCb:2015gmp, LHCb:2017smo, Aaij:2017tyk} collaborations, which show deviations from the SM predictions approximately of $3.3\sigma$ for $R_{D^{(*)}}$ and $2\sigma$ for $R_{J/\psi}$.

Many theoretical works have been done to solve this deviation. Recent studies exploring NP in the $b \to c\ell \nu$ transitions include those cited in Refs.\cite{Dassinger:2007pj, Dassinger:2008as, Feger:2010qc, Faller:2011nj, Dutta:2017xmj, Watanabe:2017mip, Alok:2017qsi, He:2017bft, Biswas:2018jun, Zhu:2018zxb, Haritha:2022ydw, Hu:2018veh, Tang:2022nqm, Mohanty:2022vyn, Haritha:2023bmv, Arslan:2023wgk, Duan:2024ayo, Deng:2025znr,S:2024adt}. Due to sizable decay width~\cite{Chang:1992pt}, it has been suggested that the $B_c \to \eta_c\ell\bar\nu_\ell$ decay channel can be another helpful platform. Reference~\cite{Qiao:2012vt} conducted an extensive study of the $B_c$-meson within nonrelativistic QCD (NRQCD)~\cite{Bodwin:1994jh} up to next-to-leading order (NLO) QCD corrections, predicting a value of $R_{\eta_c} =0.3047$. By applying the principle of maximum conformality (PMC)~\cite{Brodsky:2011ta, Brodsky:2012rj, Mojaza:2012mf, Brodsky:2013vpa} to set the effective values of the strong coupling constant $\alpha_s$ for those channels, the authors of Ref.\cite{Shen:2014msa} reported the NLO perturbative QCD (pQCD) predictions, finding that $R_{\eta_c} =0.3794$. Ref.\cite{Tran:2018kuv} utilized the covariant confined quark model (CCQM) to compute the relevant transition form factors (TFFs). These TFFs were employed to investigate the semileptonic decays of the $B_c$-meson into final charmonium states both within the SM and in scenarios beyond it, resulting in a value of $R_{\eta_c} = 0.26$. Some more approaches, such as the pQCD $k_T$-factorization approach, the three-point QCD sum rules (3PSR) and the light-front quark model (LFQM), have also been employed to investigate the semi-leptonic decays $B_c^+\to \eta_c\ell^+\nu_\ell$. Typical predictions of those approaches are $R_{\eta_c} =  0.3106$~\cite{Wang:2012lrc}, $R_{\eta_c} =  0.36$~\cite{Azizi:2019aaf}, and $R_{\eta_c}=  0.283$~\cite{Wang:2008xt}, respectively. In this study, we will examine the potential NP effects in the semi-leptonic decays of $B_c$-meson into $\eta_c$ using an effective Hamiltonian approach that incorporates all the conceivable four-fermion operators. And as an explicit explanation, we will employ two distinct methodologies to discuss possible NP effects, one focuses solely on a single Wilson coefficient, while the other considers both Wilson coefficients. Subsequently, we will analyze the implications of those NP operators on various physical observables, including the branching fraction ratios, the forward-backward asymmetries, the convexity parameter, and the polarization.

Prior to computing those physical observables, it is essential to calculate the $B_c\to \eta_c$ TFFs. We will employ the light cone sum rules (LCSR)~\cite{Braun:1988qv, Balitsky:1989ry, Chernyak:1990ag, Ball:1991bs} to compute those TFFs. The LCSR approach is an efficient tool for determining the nonperturbative parameters of hadronic states. The amplitude of the process can be factorized into a perturbatively calculable short-distance part and the nonperturbative long-distance part, which can be distributed into light cone distribution amplitudes (LCDAs). Utilizing the LCSR approach involves performing an operator product expansion (OPE) in the vicinity of the light-cone, where ${x^2} \approx 0$. The resulting nonperturbative hadronic matrix elements are then parametrized using LCDAs of various twists. Refs.\cite{Bell:2008er,Bell:2024bxg} study the LCDAs of $B$ mesons and $B_c \to \eta_c$ transitions. In this study, we primarily focus on the contribution of the $\eta_c$-meson LCDAs. Regarding the twist-2 LCDA $\phi _{2;\eta_c}$, we adopt the following Gegenbauer expansion
\begin{eqnarray}
\phi_{2;\eta_c}(x) &=& 6x\bar{x} \left(1 + {a_2}C_2^{3/2}(\xi) + {a_4}C_4^{3/2}(\xi)+\cdots \right),
\label{twist2phi}
\end{eqnarray}
where $\bar{x}=1-x$, $\xi=x-\bar{x}$, $a_n$ are $n_{\rm th}$-order Gegenbauer moments. The LCDA and its Gegenbauer moments are scale dependent, whose magnitude at any scale can be derived by using proper evolution equations from their given values at some initial scale. Throughout the manuscript, if not specially stated, we will omit the scale-dependence of those parameters in the formulas for convenience. Another type of frequently adopted moments are $\xi$-moments, and the $n_{\rm th}$-order $\xi$-moment is defined as
\begin{equation}
\langle\xi_{2;\eta_c}^{n}\rangle  = \int^1_0 dx \xi^n \phi_{2; \eta_c}(x). \label{moment_n}
\end{equation}
In this paper, we will first adopt the approach of Shifman-Vainshtein-Zakharov (SVZ) sum rules~\cite{Shifman:1978bx, Shifman:1978by} under background field theory (BFTSR) to calculate the $\xi$-moments $\langle\xi_{2;\eta_c}^{n}\rangle$, and then fix the Gegenbauer moments of $\phi_{2;\eta_c}$ via their inner relations to the $\xi$-moments as will be shown later. The BFTSR approach has been adopted to compute the twist-2 and twist-3 LCDAs of various mesons, such as the scalar mesons~\cite{Han:2013zg}, the pseudoscalar mesons~\cite{Zhong:2014jla, Zhong:2014fma, Zhong:2016kuv, Zhong:2011jf, Huang:2004tp, Huang:2005av, Zhong:2011rg, Zhang:2017rwz, Zhang:2021wnv}, the vector mesons~\cite{Fu:2016yzx, Fu:2018vap}, the axial vector mesons~\cite{Hu:2021lkl}, and etc. Those works show that the BFTSR approach offers a viable way to incorporate the nonperturbative effects and to provide a systematic description of vacuum condensates from a field-theoretic perspective. Furthermore, the QCD background field theory (BFT)~\cite{Govaerts:1983ka, Govaerts:1984bk, Huang:1989gv} provides a clear physical picture of both perturbative and nonperturbative QCD properties and establishes a systematic framework for deriving the QCD SVZ sum rules relevant to hadron phenomenology. Consequently, the BFT is a highly valuable method for computing meson distribution amplitudes.

The remaining parts of the paper are organized as follows. In Sec.~\ref{Section:II}, we present the calculation technology for the $\xi$-moments of the $\eta_c$-meson twist-2 LCDA $\phi_{2;\eta_c}$, the $B_c \to \eta_c$ TFFs, the branching ratios, the forward-backward asymmetry $\mathcal{A}_{\rm FB}({q^2})$, the convexity parameter $\mathcal{C}_F^\tau ({q^2})$, and the longitudinal and transverse polarizations $\mathcal{P}_L ({q^2})$ and $\mathcal{P}_T ({q^2})$, respectively. In Sec.~\ref{Section:III}, we give the numerical results and discussions. A discussion of possible NP effects is also given there. Section~\ref{Section:IV} is reserved for a summary.  \\

\section{Calculation Technology}\label{Section:II}

\subsection{Decay processes for $B_c \to \eta_c \ell\bar\nu_\ell$}

As we know, general effective Lagrangian for the semileptonic decay processes $B_c \to\eta_c \ell\bar\nu_\ell$ $\ell = (e,\mu,\tau)$ at quark level can be written as
\begin{eqnarray}
{\cal L}_{\rm eff} = \frac{G_F V_{cb}}{\sqrt 2}& \big[&
 (1+ V_L){\cal O}_{V_L} + V_R{\cal O}_{V_R} + S_L {\cal O}_{S_L} \nonumber \\
&& + S_R {\cal O}_{S_R}+ T_L {\cal O}_{T_L} \big], \label{Eq:Lag}
\end{eqnarray}
where the four-Fermi operators have the definitions
\begin{align}
&\mathcal{O}_{S_{L,R}} = [\bar{c} (1\mp \gamma_5) b)][\bar\ell(1-\gamma_5)\nu_\ell ],
\nonumber \\
&\mathcal{O}_{V_{L,R}} = [\bar{c}\gamma^\mu (1\mp \gamma_5) b ][\bar\ell\gamma_\mu(1-\gamma_5)\bar\nu_\ell ],
\nonumber \\
&\mathcal{O}_{T_L} = [\bar{c}\sigma^{\mu\nu} (1-\gamma_5)b] [\bar\ell\sigma_{\mu\nu}(1-\gamma_5)\nu_\ell ].
\end{align}
The $G_{\rm F}$ and $V_{cb}$ are fermi constant and CKM matrix element, respectively. In the SM, the coefficients $V_{L,R}, S_{L,R}$ and $T_L$ tend to zero, indicating that they have no contributions to the Lagrangian. On the contrary, the corrections to the SM are assumed to be generated by new physics that enter at a much higher energy scale, whose strengths at the SM scale are governed by the unknown Wilson coefficients $V_{L,R}$, $S_{L,R}$ and $T_L$, which are complex in general. Normally, the NP effects often contribute to the case of $\tau$-lepton pair due to the lepton flavor violation. Then the matrix element of $B_c \to  \eta_c \ell \bar\nu_\ell$ has the following general form,
\begin{align}
\mathcal{M} = \frac{G_F V_{cb}}{\sqrt2}&\Big\{ (1+V_L+V_R)\langle \eta_c|\bar c \gamma^\mu b|\bar B_c\rangle (\bar\ell\gamma_\mu \Gamma_L \nu_\ell)
\nonumber\\[-1ex]
& + (V_R-V_L) \langle \eta_c|\bar c\gamma^\mu \gamma_5 b|\bar B_c\rangle (\bar\ell\gamma_\mu \Gamma_L \nu_\ell)
\nonumber\\[0.4ex]
&  + (S_R+S_L) \langle \eta_c|\bar c b |\bar B_c\rangle (\bar\ell \Gamma_L \nu_\ell)
\nonumber\\[0.4ex]
& +(S_R-S_L) \langle \eta_c|\bar c\gamma_5 b |\bar B_c\rangle (\bar\ell \Gamma_L \nu_\ell)
\nonumber\\
& + T_L \langle \eta_c| \bar c\sigma^{\mu\nu} \Gamma_L  b|{\bar{B}_c} \rangle (\bar\ell\sigma^{\mu\nu} \Gamma_L \nu_\ell) \Big\},
\label{Eq:BcMccMatrix}
\end{align}
where $\Gamma_L = (1-\gamma_5)$. It is noted that the axial and pseudoscalar hadronic currents do not contribute to $B_c \to \eta_c$ decay, which leads to $V_R-V_L = 0$ and $S_R-S_L = 0$. We henceforth use the shorthand definition $S_R+S_L =S$ and $S_R-S_L = P$ throughout the body of the text.

Using Eq.~\eqref{Eq:BcMccMatrix}, the differential decay width for the semileptonic decay $B_c \to \eta_c \ell\bar\nu_\ell$ versus the squared momentum transfer ($q^2$) of the $(\ell \nu_\ell)$-pair and $\cos\theta_\ell$ with $\theta_\ell$ being the lepton polar angle can be calculated within the helicity amplitude approach
\begin{align}
\frac{d^2\Gamma(B_c \to \eta_c \ell\bar\nu_\ell)}{dq^2d\cos\theta_\ell} = \frac{\cal G}{64(2\pi)^3} H_{\mu\nu}L^{\mu\nu} (\theta_\ell),
\end{align}
where ${\cal G} = G_F^2|V_{\rm cb}|^2|{\bf p_2}|v^2 / m_{B_c}^2$ and the Fermi constant $G_F=1.166\times10^{-5}\;{\rm GeV}^{-2}$. $|{\bf p_2}| = \lambda^{1/2}/2m_{B_c}$ is the momentum of $\eta_c$ in the $B_c$-meson rest frame, where the phase-space factor $\lambda = (m_{B_c}^2+m_{\eta_c}^2-q^2)^2 -4m_{B_c}^2m_{\eta_c}^2$. $v= (1- m_\ell^2/q^2)$ is the lepton velocity in the $(\ell \bar{\nu}_\ell)$-pair center-of-mass frame and $H_{\mu\nu}L^{\mu\nu}$ is the contraction of the hadronic and the leptonic tensors. The advantage of helicity amplitude approach is that the helicity amplitude corresponds to a transition amplitude with deﬁnite spin-parity quantum number in the lepton pair center-of-mass frame. The helicity amplitudes to calculate the angular distribution in the presence of new physics operators for the semileptonic decays considered here can be found in Refs.\cite{Ivanov:2016qtw, Cohen:2018vhw}. More explicitly, the differential decay distribution for $B_c\to \eta_c \ell\bar\nu_\ell$ can be written as
\begin{align}
&\frac{d^2\Gamma(B_c\to\eta_c\ell\bar\nu_\ell)}{dq^2d\cos\theta_\ell} = \frac{{\cal G}q^2}{16(2\pi)^3} \Big\{|1\!+\!V_L \!+\! V_R|^2 (|H_0|^2 \sin^2\theta_\ell
\nonumber\\
&\qquad +2\delta_\ell \, |H_t-H_0\cos\theta_\ell|^2) \,+\, |S|^2|H_P^S|^2
+ 16|T_L|^2\big(2\delta_\ell
\nonumber\\[0.5ex]
&\qquad + ( 1-2\delta_\ell)\cos^2\theta_\ell \big)\,|H_T|^2 + 2\sqrt{2\delta_\ell}({\rm Re}S + S V_L) H_P^S
\nonumber\\[0.5ex]
&\qquad \times \big( H_t\,- \, H_0\cos\theta_\ell \big) \,+\, 8\, \sqrt{2\delta_\ell} \,({\rm Re}T_L \,+\, T_L V_L ) \, \big( H_0
\nonumber\\[0.5ex]
&\qquad-H_t\cos\theta_\ell \big) H_T -8 H^S_P H_T \cos\theta_\ell T_L S \Big\},
\label{Eq:dGamma-Bcetac}
\end{align}
where the helicity flip-factor $\delta_\ell = m_\ell^2/{2q^2}$. The multiplication rule between different complex Wilson coefficients are $(C_1 C_2)   = {\rm Re}C_1~ {\rm Re} C_2 + {\rm Im} C_1\,{\rm Im} C_2$, $(C_1 C_2)^* = {\rm Im}C_1\,{\rm Re} C_2 - {\rm Im} C_1\,{\rm Re} C_2$, where $(C_1, C_2)\in [V_{L,R},S_{L,R},T_L,S,P]$. The hadronic helicity form factors $H_a$ can be written in terms of the traditional TFFs. For the $B_c\to \eta_c \ell\bar\nu_\ell$ decay processes, {\it e.g.} the four helicity form factors $H_a = H_t, H_0, H_P^S, H_T$ are related to the TFFs $f_{+,0,T}^{B_c\to \eta_c}(q^2)$ via the following way~\cite{Leljak:2019eyw}
\begin{align}
&H_t = (m_{B_c}^2-m_{\eta_c}^2)/\sqrt{q^2}f_0(q^2),\label{Eq:HFFs1_etac}  \\
&H_0 = (\lambda/{q^2})^{1/2} f_+(q^2),\label{Eq:HFFs2_etac}  \\
&H_P^S = (m_{B_c}^2-m_{\eta_c}^2)/(m_b-m_c) f_0(q^2),\label{Eq:HFFs3_etac}  \\
&H_T = \sqrt{\lambda} / (m_{B_c}+m_{\eta_c}) f_T(q^2).\label{Eq:HFFs4_etac}
\end{align}

The differential distribution (\ref{Eq:dGamma-Bcetac}) can be rearranged as a more compact form as a power function of $\cos \theta_\ell$, {\it e.g.}
\begin{eqnarray}
\frac{d\Gamma}{dq^2 d\cos\theta_\ell} &=& \frac{{\cal G}q^2}{256\pi^3} \bigg[A_{\eta_c}(q^2) + B_{\eta_c}(q^2) \cos\theta_\ell \nonumber\\
&&\quad\quad\quad\quad + C_{\eta_c}(q^2) \cos^2\theta_\ell\bigg],  \label{eq:dGABC}
\end{eqnarray}
where the expressions of the functions $A_{\eta_c}(q^2)$, $B_{\eta_c}(q^2)$, and $C_{\eta_c}(q^2)$ are
\begin{align}
A_{\eta_c}(q^2) &= |1 +  V_L + V_R|^2(|H_0|^2 + 2\delta_\ell|H_t|^2) + |S|^2
\nonumber\\[0.4ex]
&\times |H_P^S|^2+ 32\delta_\ell |T_L|^2|H_T|^2\,+2\sqrt{2\delta_\ell}\Big[({\rm Re} S
\nonumber \\
& + S V_L) H_P^S H_T + 4 ({\rm Re} T_L + T_L V_L)H_0 H_T\Big],
\\
B_{\eta_c}(q^2) &= -2\sqrt{2\delta_\ell}\, \Big[\sqrt{2\delta_\ell} \,|1  \,+   V_L \,+  V_R|^2  |H_t||H_0|
\nonumber\\[0.4ex]
& + ({\rm Re} S + SV_L)\,H_P^S H_0  \,+  4 ({\rm Re} T_L + T_L V_L)
\nonumber\\[0.4ex]
& \times H_t H_T \Big] - 8 (T_L S) H_P^SH_T, \\
C_{\eta_c}(q^2) &= \, |1\,+\,\,V_L\,+\,V_R|^2 \,(2\delta_\ell \,- \,1)\,|\,H_0\,|^2 \,+\, 16
\nonumber\\[0.4ex]
&\times |T_L|^2(1 - 2\delta _\ell)|H_T|^2.
\end{align}
We then define two $\theta_\ell$-dependent observables, e.g., the forward-backward lepton asymmetry ${\mathcal A}_{\rm FB}^{\eta_c}$ and the convexity parameter ${\mathcal C}^{\tau, \eta_c}_{\rm F}$, which are
\begin{align}
& {\mathcal A}_{\rm FB}^{\eta_c}(q^2)  =  \frac{3 B_{\eta_c}(q^2)}{2[3A_{\eta_c}(q^2)+ C_{\eta_c}(q^2)]},
\\[2ex]
& {\mathcal C}^{\tau, \eta_c}_{\rm F}(q^2) = \frac{3 C_{\eta_c}(q^2)}{3A_{\eta_c}(q^2)+ C_{\eta_c}(q^2)}.
\end{align}

Moreover, the polarization of the emitted $\tau$ in the $W^-$ rest frame is helpful for testing NP operators. The differential decay rate for a given spin projection in a given direction can be derived with the inclusion of the spin projection operator such as $(1+{\gamma_5}{\slash\!\!\! s_i})/2$ with $i=L$ or $T$, where the longitudinal polarization vector $s_L$ and the transverse polarization vector $s_T$ of $\tau^-$ in $W^-$ rest frame are~\cite{Faessler:2002ut, Gutsche:2013oea, Gutsche:2015mxa}, $s_L =(|\vec{p}_\tau|, E_\tau\sin\theta_\tau,0,E_\tau\cos\theta_\tau)/{m_\tau}$ and $s_T =(0, \cos\theta_\tau, 0, -\sin\theta_\tau)$, respectively. The longitudinal and the transverse polarization components of $\tau$ are then defined as
\begin{align}
\mathcal P_{\rm L,T}^{\eta_c} (q^2) &= \frac{3 P_{\rm L,T}^{\eta_c}(q^2)}{2[3A_{\eta_c}(q^2)+ C_{\eta_c}(q^2)]},
\end{align}
where the longitudinal and transverse polarizations are
\begin{align}
P_L^{\eta_c}(q^2) &= |1+ V_L+ V_R|^2 \big[3\delta_\tau|H_t|^2-(1  - \! \delta_\tau)|H_0|^2 \big]~~~
\nonumber \\
& + 3\sqrt{2\delta_\tau}({\rm Re}S \! + \! SV_L ) (H_P^S H_t) \!+\! \frac32|S|^2|H_P^S|^2
\nonumber \\
& +8\, (1\, - \, 4\delta_\tau) \, |T_L|^2 \,|H_T|^2 \,-\, 4\,\sqrt{2\delta_\tau}\,({\rm Re}T_L
\nonumber \\[1.2ex]
&+ T_LV_L ) (H_0 H_T),
\\
P_T^{\eta_c}(q^2) &= \frac{3\pi\sqrt{\delta_\tau}}{2\sqrt{2}}\Big[|1 \,+ V_L\,+V_R|^2  (H_0 H_t) +\frac1{\sqrt{2\,\delta_\tau}}
\nonumber \\
&\times ({\rm Re}S \, + \, SV_L)\,(H_P^SH_0) \, +\,  4\,\sqrt{2\,\delta_\tau}\,({\rm Re}T_L
\nonumber \\[0.8ex]
&+ T_L V_L)(H_t H_T)+ 4T_L S (H_P^S H_T)  \Big].
\end{align}

\subsection{The $B_c \to \eta_c$ TFFs within the QCD LCSR}
As shown above, we need to know four helicity form factors $H_t$, $H_0$, $H_P^S$, and $H_T$ for the transition of $B_c \to \eta_c$. They are related to the traditional TFFs $f_{+,0,T}^{B_c\to \eta_c}(q^2)$ that can be calculated by using the following matrix elements:
\begin{align}
&\langle \eta _c(\tilde{p})|\bar{c}\gamma _{\mu}b|B_c(p)\rangle =\left[ (\tilde{p}+p)_{\mu}-\frac{m_{B_c}^{2}-m_{\eta _c}^{2}}{q^2}q_{\mu} \right] f_+(q^2)
\nonumber\\
&\hspace{2.8cm}+\left[ \frac{m_{B_c}^{2}-m_{\eta _c}^{2}}{q^2}q_{\mu} \right] f_0(q^2),
\\
&\langle \eta _c(\tilde{p})|\bar{c}\sigma _{\mu \nu}q^{\nu}b|B_c(p)=\frac{i}{m_{B_c}+m_{\eta _c}}\bigg[ q^2(\tilde{p}+p)_{\mu}-(m_{B_c}^{2}
\nonumber\\
&\hspace{3.2cm}-m_{\eta _c}^{2})q_{\mu}\bigg] f_T(q^2) ,
\end{align}
where $q=p-{\tilde p}$ is the momentum transfer, and it varies within the range $0\leq q^2\leq (m_{B_c}-m_{\eta_c})^2$. In the large recoil region, we have $f_+(0)=f_0(0)$. The tensor TFF satisfies the following relation
\begin{gather}
{f_T}({q^2}) = \frac{{m_{{B_c}}^3}}{{{m_b}(m_{{B_c}}^2 - {m^2_{{\eta _c}}})}}{f_ + }({q^2}).\label{Eq:fT}
\end{gather}

These TFFs can be derived from the correlation function (correlator) of the $T$-product between the weak current and the interpolating current of the $B_c$-meson, taken between the vacuum and an external on-shell $\eta_c$-meson state, i.e.
\begin{align}
\Pi(p, q) =
i\int d^4x e^{iq \cdot x} \langle \eta_c (p)|T\{j_{\eta_c}^A (x), j_{B_c}^\dagger(0) \} |0\rangle,
\label{eq:correlator}
\end{align}
where the current $j_{\eta_c}^A (x) = \bar c(x)\gamma_\mu\gamma_5 b(x)$, and the current $j_{B_c}^\dagger(x)=i\bar{b}(x)\gamma_5 c(x)$, which has the same quantum state as the pseudoscalar $B_c$-meson with $J^P=0^{-}$.

The correlator (\ref{eq:correlator}) is an analytic function of $q^2$, and it is defined in both the spacelike and timelike $q^2$-regions. In timelike $q^2$-region, it can be treated by inserting a complete set of intermediate hadronic states in physical region and obtain its hadronic representation by isolating out the pole term of the lowest pseudoscalar $B_c$-meson, e.g the ground state can be isolated by using $\langle B_c |j^{\dagger}_{B_c} | 0 \rangle = m_{B_c}^2 f_{B_c} /(m_b + m_c)$. The contributions from higher resonances and continuum states above the threshold $s_0^{B_c}$ can be expressed using the dispersion integration. The spectral density associated with this integration can be approximated through the quark-hadron duality ansatz~\cite{Shifman:1978bx, Shifman:1978by}. While in spacelike $q^2$-region, the correlator~(\ref{eq:correlator}) can be expanded by the operator product expansion, which can be preprocessed by contracting the $b$-quark operator to a free propagator,
\begin{align}
&\langle0|b_\alpha^i(x)\bar b_\beta^j(0)|0\rangle = -i\int \frac{d^4k}{(2\pi)^4}e^{-ik\cdot x}\bigg\{\delta^{ij}\frac{\DS k + m_b}{m_b^2-k^2}
\nonumber\\
&\qquad  +g_s\int_0^1 dv G^{\mu\nu}(vx)\left(\frac{\lambda}{2}\right)^{ij} \bigg[\frac{\DS k+m_b}{2(m_b^2 - k^2)^2}\sigma_{\mu\nu}
\nonumber\\
&\qquad + \frac1{m_b^2-k^2}vx_\mu\gamma_\nu\bigg]\bigg\}_{\alpha\beta}.
\end{align}
The correlator~(\ref{eq:correlator}) can be decomposed into the product of perturbatively calculable hard-scattering kernels and nonperturbative, universal LCDAs, ordered by increasing twist~\cite{Ball:2004ye}.

Following the standard LCSR procedures for the heavy-to-light TFFs, by equating the correlator in different regions and applying the conventional Borel transformation to suppress the contributions from the unknown continuum states and higher resonances, our final LCSRs for the $B_c \to \eta_c$ TFFs are
\begin{widetext}
\begin{align}
f^{B_c\to\eta_c}_+(q^2)&= \frac{f_{\eta_c} m_b}{f_{B_c}m_{B_c}^2} \! \int_0^1\! d u e^{(m_{B_c}^2 - s)/M^2} \bigg\{ m_b \bigg[ \frac{1}{2u} \Theta (c(u,s_0))\phi_{3;\eta_c}^p(u) - \!\frac{2m_b^2}{u^3M^4} \widetilde{\widetilde\Theta}(c(u,s_0))\psi_{4;\eta_c}(u)\hspace{2cm}
\nonumber
\\
& + \frac{1}{u M^2} \widetilde\Theta (c(u,s_0))\phi_{4;\eta_c}(u)+ \frac{2 m_b^2}{u^3 M^4} \widetilde{\widetilde\Theta}(c(u,s_0))\Phi_{4I;\eta_c}(u) \bigg] +\frac{m_{\eta_c}^2}{2(m_u + m_b)}\bigg[\Theta (c(u,s_0))
\nonumber
\\
&\times\phi_{2;\eta_c}(u) \,+\, \bigg(\frac{m_b^2 + q^2}{6 u^2 M^2}\widetilde\Theta (c(u,s_0))
+\frac{1}{3u}\Theta (c(u,s_0))\bigg)\times~\phi_{3;\eta_c}^\sigma(u) \bigg] \,-\,\frac{f_{3\eta_c}}{f_{\eta_c} }\times\frac{I_{3\eta_c}(u)}{u}\,
\nonumber
\\
& - \frac12\frac{m_b}{m_b^2-q^2}I_{4\eta_c}(u) \bigg\}+ \frac{\alpha_s C_F}{8\pi m_{B_c}^2 f_{B_c}}F_1(q^2,M^2,s_0),
\label{eq:fp}
\\[1ex]
f^{B_c\to\eta_c}_0(q^2)& = \frac{f_{\eta_c} m_b}{f_{B_c} m_{B_c}^2(m_{B_c}^2-m_{\eta_c}^2)}\int_0^1 du e^{(m_{B_c}^2 - s)/M^2} \bigg\{{\cal Q}_+ \, m_b \bigg[ \frac{1}{2u}\, \Theta (c(u,s_0))\,\phi_{3;\eta_c}^p(u)- \frac{2m_b^2}{u^3 M^4}
\nonumber
\\
&\times\widetilde{\widetilde\Theta}(c(u,s_0))\psi_{4I;\eta_c}(u)
\! + \! \frac{1}{u M^2}\widetilde\Theta (c(u,s_0))\phi_{4;\eta_c}(u) + \frac{2 m_b^2}{u^3 M^4} \widetilde{\widetilde\Theta}(c(u,s_0))\Phi_{4;\eta_c}(u)\bigg]+{\cal Q}_+
\nonumber
\\
&\times\frac{m_{\eta_c}^2}{2(m_c + m_b)}\bigg[\Theta (c(u,s_0))\phi_{2;\eta_c}(u) \!+\! \bigg(\frac{1}{3u}
\Theta(c(u,s_0)) \!+\! \frac{m_b^2 + q^2}{6 u^2 M^2}\widetilde\Theta (c(u,s_0))\bigg)\phi_{3;\eta_c}^\sigma(u) \bigg]
\nonumber
\\
& + q^2~\bigg[\frac{4 m_b}{u^2 M^2}~\widetilde\Theta (c(u,s_0))~\phi_{4;\eta_c}(u)~+~ \frac{2 m_{\eta_c}^2}{u(m_c + m_b)}~\Theta (c(u,s_0))~ \phi_{2;\eta_c}(u)\bigg]\bigg\}~-~\int_0^1 du
\nonumber
\\
&\times e^{(m_{B_c}^2 - s)/M^2} \frac{{\cal Q}_-}{m_{B_c}^2 - m_{\eta_c}^2} \bigg[\frac{m_b f_{3\eta_c}}{u m_{B_c}^2 f_{B_c}} I_{3;\eta_c}(u)+\frac{m_b^2 f_{\eta_c}}{2m_{B_c}^2 f_{B_c}(m_b^2-q^2)}I_{4;\eta_c}(u)\bigg]
+\frac{\alpha_s C_F}{4\pi}
\nonumber
\\
&\times\frac{1}{2m_{B_c}^2 f_{B_c}}\bigg[F_1(q^2,M^2,s_0) + \frac{q^2}{m_{B_c}^2-m_{\eta_c}^2}(\widetilde F_1(q^2,M^2,s_0) - F_1(q^2,M^2,s_0))~\bigg],\quad
\label{eq:f0}
\end{align}
\end{widetext}
where ${\cal Q}_\pm = q^2 \pm (m_{B_c}^2 - m_{\eta_c}^2)$ and $c(u,s_0)=u s_0 - m_b^2 + \bar u q^2 - u \bar u m_{\eta_c}^2$. $\Theta(c(\varrho,s_0))$ is the usual step function: when $c(\varrho,s_0)<0$, it is zero; otherwise, it is $1$. $\widetilde\Theta (c(u,s_0))$ and $\widetilde{\widetilde\Theta}(c(u,s_0))$ are defined via the integration
\begin{align}
& \int_0^1 \frac{du}{u^2 M^2} e^{-s(u)/M^2}\widetilde\Theta(c(u,s_0))f(u) \nonumber\\
& =\int_{u_0}^1\frac{du}{u^2 M^2} e^{-s(u)/M^2}f(u) + \delta(c(u_0,s_0)), \label{Theta1} \\
& \int_0^1 \frac{du}{2u^3 M^4} e^{-s(u)/M^2} \widetilde{\widetilde\Theta}(c(u,s_0))f(u) \nonumber\\
& =\int_{u_0}^1 \frac{du}{2u^3 M^4} e^{-s(u)/M^2}f(u)+\Delta(c(u_0,s_0)), \label{Theta2}
\end{align}
where $\delta(c(u,s_0)) = e^{-s_0/M^2}f(u_0)/{\cal C}_0$ and
\begin{align}
\Delta(c(u,s_0))&= e^{-s_0/M^2}\bigg[\frac{1}{2 u_0 M^2}\frac{f(u_0)} {{\cal C}_0}
\nonumber\\
&-\frac{u_0^2}{2 {\cal C}_0} \frac{d}{du}\left( \frac{f(u)}{u{\cal C}} \right) |_{u = {u_0}}\bigg],
\end{align}
${\mathcal C}_0 = m_b^2 + {u_0^2}m_{\eta_c} ^2 - {q^2}$ and $u_0\in[0,1]$ is the solution of $c(u_0,s_0)=0$. The shorthand notations introduced for the integrals over two and three-particle DA's are
\begin{align}
I_{3;\eta_c}(u) &=\frac{d}{du}\left[\int_0^u d\alpha_1 \int_{\frac{u-\alpha_1}{1-\alpha_1}}^1 dv \Phi_{3;\eta_c}(\alpha_i) \right]
\\
I_{4;\eta_c}(u)&=\frac{d}{du}\Bigg\{\int_0^u\! d\alpha_1 \int_{\frac{u-\alpha_1}{1-\alpha_1}}^1 \frac{dv}{v} \Bigg[ 2\Psi_{4;\eta_c}(\alpha_i)
\nonumber\\
&- \Phi_{4;\eta_c}(\alpha_i)+2\widetilde{\Psi}_{4;\eta_c}(\alpha_i) -\widetilde{\Phi}_{4;P}(\alpha_i)\Bigg] \Bigg\},  \label{eq:fplusBpiLCSR3part}
\end{align}
where $\alpha_2 = 1-\alpha_1-\alpha_3$ and $\alpha_3=(u-\alpha_1)/v$. The NLO terms in Eqs.~\eqref{eq:fp} and \eqref{eq:f0} have the form of the dispersion relation:
\begin{align}
F_1(q^2,M^2,s_0)  = \frac{1}{\pi}\!\int_{m_b^2}^{s_0} ds e^{(m_{B_c}^2-s)/M^2}\,{\rm Im} F_1(q^2,s)
\end{align}
with
\begin{align}
&{\rm Im} F_1(q^2,s) = f_{\eta_c}  \int_0^1 du \bigg [ {\rm Im} T_1(q^2,s,u) \phi_{2;\eta_c}(u)
\nonumber\\
&\, + {\rm Im} T_1^p(q^2,\!s,\!u)\phi_{3;\eta_c}^p(u) \! + \!{\rm Im} T_1^\sigma(q^2,\!s,\!u)\phi_{3;\eta_c}^\sigma(u)\bigg ].
\label{eq:Imconvol}
\end{align}
Here to have an idea on how the NLO QCD corrections affect the LCSRs, we have also considered the NLO corrections to the twist-2 and twist-3 parts, and the expressions of the imaginary parts of the amplitudes can refer to $B\to \pi$ process~\cite{Li:2015cta}. The NLO QCD corrections have UV divergence and infrared collinear divergence. The UV divergence can be canceled by the heavy quark mass renormalization, and the infrared collinear divergence term can be absorbed into the evolution of LCDA~\cite{Li:2015cta}. The NLO amplitudes $\widetilde{F}_1(q^2,s,u)$ have the same expression as $T_1^{(p,\sigma)} (q^2,s,u) \to \widetilde T_1^{(p,\sigma)} (q^2,s,u)$. Regarding the twist-2 LCDA $\phi_{2;\eta_c}(x)$, we adopt the truncated form of the Gegenbauer expansion as shown by Eq.~(\ref{twist2phi}). It will be found that the twist-2 LCDA $\phi_{2;\eta_c}$ provides dominant contribution to the LCSRs either directly or indirectly. Thus, as required, those TFFs can provide us a useful platform for testing the properties of $\phi_{2;\eta_c}$.

\subsection{The $\eta_c$-meson twist-2 LCDA $\phi_{2;\eta_c}$}

The QCD Lagrangian within the framework of BFT can be obtained from conventional QCD Lagrangian by replacing the gluon field ${\cal A}_\mu^A(x)$ and quark field $\psi(x)$ to the following ones $\mathcal{A}^A_\mu(x) \to \mathcal{A}^A_\mu(x) + \phi^A_\mu(x)$ and $\psi(x) \to \psi(x) + \eta(x)$. Here $\mathcal{A}^A_\mu(x)$ with $A =(1, \ldots, 8)$ and $\psi(x)$ are gluon and quark background fields. $\phi^A_\mu(x)$ and $\eta(x)$ are gluon and quark quantum fields, {\it i.e.} the quantum fluctuation on the background fields. The QCD Lagrangian within the BFT is given~\cite{Huang:1989gv}
\begin{align}
{\cal L}_{\rm QCD} = - \frac14 G_{\mu\nu}^a G^{a,\mu\nu} + i\bar\psi(x) \slash \!\!\!\!  D \psi(x) - m\bar\psi(x) \psi(x),
\end{align}
which satisfies the equations of motion
\begin{eqnarray}
(i \slash \!\!\!\!  D - m)\psi(x) &=& 0, \label{Eq:EOM1}
\\
\widetilde{D}^{AB}_\mu G^{B\nu\mu}(x) &=& g_s \bar{\psi}(x) \gamma^\nu T^A \psi(x), \label{Eq:EOM2}
\end{eqnarray}
where $D_\mu = \partial_\mu - ig_s T^A \mathcal{A}^A_\mu(x)$ and $\widetilde{D}^{AB}_\mu = \delta^{AB} - g_s f^{ABC} \mathcal{A}^C_\mu(x)$ are fundamental and adjoint representations of the gauge covariant derivative, respectively. The physical observables should be gauge independent, one may take different gauges for the quantum fluctuations and the background fields such that to make the sum rules calculation relatively simpler. Practically, we adopt the background gauge, $\widetilde{D}^{AB}_\mu \phi^{B \mu}(x) = 0$, for the gluon quantum field~\cite{Novikov:1983gd, Hubschmid:1982pa}, and the Schwinger gauge or the fixed-point gauge, $x^\mu \mathcal{A}^A_\mu(x) = 0$, for the background field~\cite{Shifman:1980ui}. Using those inputs, the quark propagator $S_F(x,0)$ and the vertex operators $\Gamma (z\cdot \tensor{D})^n$ are ready to be derived, whose explicit expressions up to dimension-six operators have been given in Ref.\cite{Zhong:2014jla}.

The $\eta_c$-meson twist-2 LCDA, {\it i.e.} $\phi_{2;\eta_c}$ is defined via the following equation
\begin{eqnarray}
\langle 0|\bar c(z)\DS z \gamma_5[z,-z] c (-z) |\eta_c(\tilde p) \rangle =&
\nonumber\\
i (z\cdot \tilde p) f_{\eta_c}  \int^1_0 dx & e^{i\xi(z\cdot \tilde p)} \phi_{2;\eta_c}(x),
\label{Eq:DAdefinition1}
\end{eqnarray}
where $[z,-z] = P \exp[ig\int^z_{-z} dx^\mu A_\mu (x)]$ and $\xi = (2x-1)$. Expanding the left-hand-side of Eq.~(\ref{Eq:DAdefinition1}) near the light-cone $z^2 \rightsquigarrow 0$ and writing the exponent in the right-hand side of Eq.~(\ref{Eq:DAdefinition1}) as a power series, we then obtain the definition of the LCDA $\xi$-moments, in which $f_{\eta_c}$ is $\eta_c$-meson decay constant, {\it e.g.,}
\begin{align}
&\langle 0|\bar c(0)\DS z\gamma_5(iz\cdot \tensor{D})^n c(0) |\eta_c(\tilde p)\rangle = i(z \cdot \tilde p)^{n+1} f_{\eta_c} \langle \xi_{2;\eta_c}^n\rangle,
\label{moment_definition}
\end{align}
where $q$ is momentum, $(z\cdot \tensor{D})^n = (z\cdot \overrightarrow{D} - z\cdot \overleftarrow{D})^n$. As a special case of $\xi$-moments (\ref{moment_n}), the $0_{\rm th}$-moment satisfies the normalization condition
\begin{eqnarray}
& \langle \xi_{2;\eta_c}^0\rangle = \int_0^1 dx \phi_{2;\eta_c}(x) = 1,  \label{moment_0}
\end{eqnarray}
Setting $n=0$ in Eq.~(\ref{moment_definition}), we obtain
\begin{eqnarray}
&\langle 0| \bar c(0) \DS z \gamma_5 c(0) |\eta_c (\tilde p)\rangle  = i(z \cdot \tilde p) f_{\eta_c}.
\label{moment_definition_0}
\end{eqnarray}

To derive QCD sum rules for the moments of $\phi_{2;\eta_c}(x)$, we consider the following correlator:
\begin{eqnarray}
\Pi_{2;\eta_c} &=&  i \int d^4x e^{iq\cdot x} \langle 0| T \{ J_n^{\eta_c}(x), J^{\eta_c \dag}_0(0) \} |0\rangle
\nonumber\\
&=& (z \cdot \tilde p)^{n+2} I (q^2), \label{Eq:correlatoretac}
\end{eqnarray}
where $z^2 \rightsquigarrow 0$, and the two currents $J_n^{\eta_c}(x) = \bar c(0) \DS z \gamma_5 (i z\cdot \tensor{D})^n c(x)$ and $J^{\eta_c \dag}_0(0)  = \bar c(0) \DS z \gamma_5 c(0)$.

In physical region, the correlator~\eqref{Eq:correlatoretac} can be quantified by inserting a complete set of the intermediate hadronic states into the matrix element. With the help of Eq.~(\ref{moment_definition}), we obtain
\begin{eqnarray}
\frac{1}{\pi} \textrm{Im} I^{(n,0)}_{\rm had}(q^2) &=\delta (q^2 - m_{\eta_c}^2) f_{\eta_c}^2 \langle\xi_{2;\eta_c}^n\rangle
\nonumber\\
& + \rho ^{\rm cont}(q^2) \theta (q^2 - s_{\eta_c}),
\end{eqnarray}
where $s_{\eta_c}$ is the continuum threshold parameter, $\theta$ is the usual step function, and $\rho ^{\rm cont}$ stands for the hadron spectrum density from the continuum states. Using the quark-hadron duality, $\rho ^{\rm cont}$ can be written as
\begin{align}
\rho ^{\rm cont}(s) = \frac{1}{\pi}{\rm Im} I_{\rm pert}(s).
\end{align}
In deep Euclidean region $q^2 < 0$, one can apply the operator product expansion for the correlator \eqref{Eq:correlatoretac}. As a combination of the correlator within the different $q^2$-regions, the sum rules for $\langle\xi_{2;\eta_c}^n\rangle$ can be derived by using the dispersion relation. As a final step, we apply the Borel transformation to suppress contributions from excited states, continuum states, and high-dimensional operators. And our final QCD sum rules for the moments of $\eta_c$ leading-twist LCDA can be written as
\begin{align}
\langle\xi_{2;\eta_c}^n\rangle & = \frac{M^2}{f_{\eta_c}^2}e^{m_{\eta_c}^2/M^2} \bigg\{\frac{1}{\pi M^2} \int^{s_{\eta_c}}_{t_{\rm min}} ds e^{-s/M^2} \textrm{Im} I_{\rm pert}(s)
\nonumber\\
& + \hat{\cal B}_{M^2} I_{\langle G^2\rangle }(Q^2) + {\cal B}_{M^2} I_{\langle G^3\rangle }(Q^2)\bigg\},
\label{srborel}
\end{align}
where
\begin{widetext}
\begin{align}
& {\rm Im} I_{\rm pert}(s) = \frac{3}{8~\pi~(n+1)~(n+3)} \left\{ ~v^{n+1} \left[~(n+1)~ \frac{v^2-1}{2} \,-\, 1\,\right] \,-\, (v\,\to -v) \right\},
\\[2ex]
&\hat{\cal B}_{M^2} I_{\langle G^2\rangle }(Q^2)  = \frac{\langle \alpha_s G^2\rangle }{6\pi}~ \int^1_0~ dx ~e^{-m_c^2/(x\bar xM^2)} \bigg\{\bigg[\frac{1}{2} \xi ^n x^2\bar x ^2 + n(n-1) \xi^{n-2} x^3\bar x^3 \bigg] \frac{1}{M^4 x^2 \bar x ^2}\xi^n m_c^2 x\bar x\frac{x^3+\bar x ^3}{2M^6x^3\bar x ^3} \bigg\},
\\[2ex]
&{\cal B}_{M^2} I_{\langle G^3\rangle }(Q^2)  = \frac{\langle g_s^3 f G^3\rangle }{(4\pi)^2}\int^1_0~ dx~ e^{-m_c^2/(x\bar x M^2)} \bigg\{ \bigg[-\frac{45}{8} \xi^n x\bar x (x^3+\bar x^3)-\xi^n x^2\bar x^2 \bigg(\frac{16n}{9}x\bar x + \frac{22n+69}{72}\bigg) - \xi ^{n-2}
\nonumber\\
&\hspace{2.3cm}
\times x^3\bar x ^3 \frac{n(n-1)}{9} ((n+1) x\bar x  + 16x^2+16\bar x^2) \bigg] \frac{1}{2M^6x^3\bar x^3}+\bigg[-\frac{3}{4}\xi^n x\bar x m_c^2(x^4+\bar x^4)+ \frac{11n}{6}m_c^2\xi^{n-1}
\nonumber\\
&\hspace{2.3cm}
\times x^2\bar x^2 (x^3-\bar x^3)- \frac{n(n-1)}{3} \xi^{n-2}x^4\bar x^4 m_c^2 + \xi^n m_c^2 x^2\bar x^2 \bigg[ -\frac{23}{12}(x^2 + \bar x^2) +\frac{1}{3} x\bar x  + 2 \bigg] -\frac{8n}{3} m_c^2 \xi^n
\nonumber\\
&\hspace{2.3cm}
\times x^3\bar x^3 \bigg]\frac{1}{6M^8x^4\bar x^4}+ \frac25 \xi ^n x\bar x m_c^4 \frac{x^5 + \bar x ^5}{24M^{10}x^5\bar x ^5} \bigg\}.
\end{align}
\begin{figure}[t]
\centering
\includegraphics[width=0.42\textwidth]{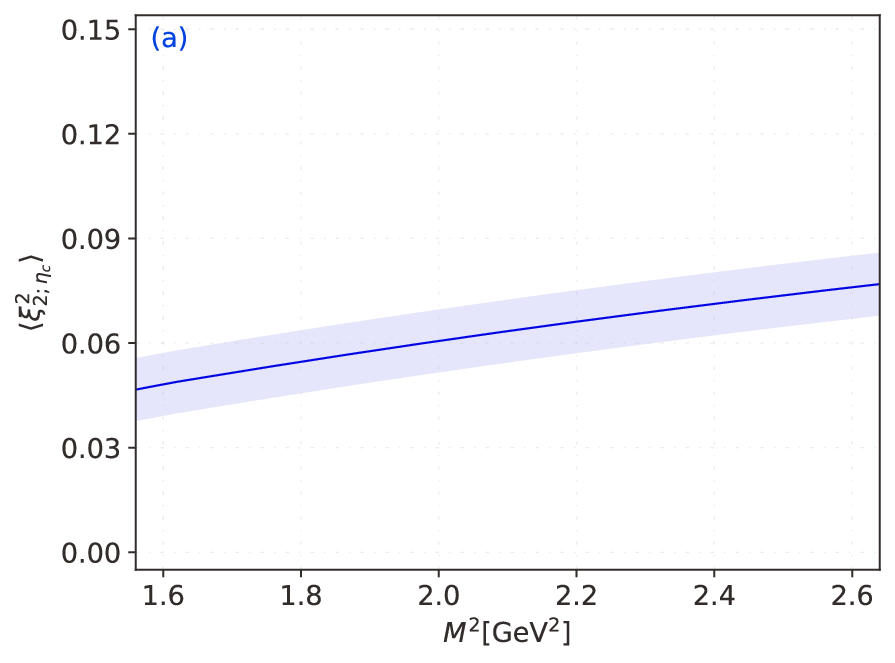}\includegraphics[width=0.42\textwidth]{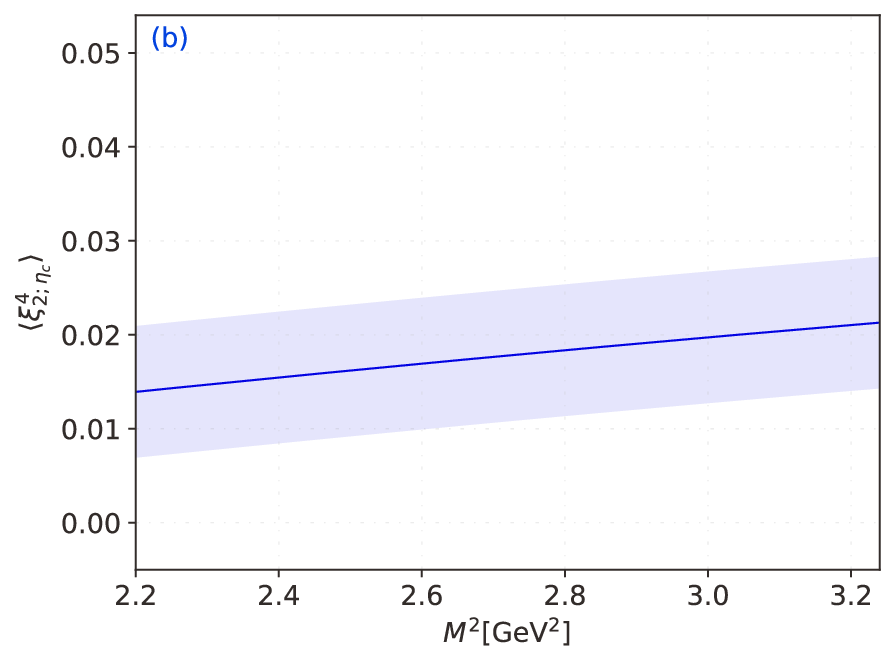}
\caption{The moments of the twist-2 LCDA $\langle \xi^n_{2;{\eta_c}} \rangle$ for $\mu=\sqrt{M^2}$ versus the Borel parameter $M^2$. The solid lines show their central values, and the shaded bands represent their uncertainties. Among them, figure (a) is the $\langle \xi^2_{2;{\eta_c}} \rangle$, figure (b) is the $\langle \xi^4_{2;{\eta_c}} \rangle$.} \label{Fig:xiM2}
\end{figure}
\end{widetext}
Using the $\langle\xi_{2;\eta_c}^n\rangle$-moments, we are able to derive the Gegenbauer moments $a_n^{\eta_c}$ via their relations to these $\xi$-moments. And the relations for the first two ones are
\begin{eqnarray}
\langle\xi_{2;\eta_c}^2\rangle &=& \frac{1}{5}  + \frac{12}{35} a_2^{\eta_c}, \label{rel_mom2}\\
\langle\xi_{2;\eta_c}^4\rangle &=& \frac{3}{35} + \frac{8}{35}  a_2^{\eta_c} + \frac{8}{77} a_4^{\eta_c}.
\label{rel_mom4}
\end{eqnarray}

In the above formulas, we have omitted the scale dependence of the LCDA and their moments for convenience. The Gegenbauer moments $a_n^{\eta_c}$ are usually known at an initial scale $\mu_0$ at the order of $\Lambda_{\rm QCD}$, which can be evolved to any scale via proper evolutions, {\it e.g.}
\begin{align}
a^{\eta_c}_n(\mu) = \left( \frac{\alpha_s(\mu)}{\alpha_s(\mu_0)} \right)^{\frac{\epsilon_n}{4\pi b_0}} a^{\eta_c}_n(\mu_0),
\label{SEE}
\end{align}
where
\begin{align}
\epsilon_n =\frac{4}{3} \left( 1 - \frac{2}{(n+1)(n+2)} + 4 \sum^{n+1}_{j=2} \frac{1}{j} \right).
\end{align}
For the running coupling, we adopt
\begin{align}
\alpha_s(\mu) = \frac{1}{b_0 t} \left[ 1 - \!\frac{b_1}{b_0^2} \frac{\ln t}{t} +\!\frac{b_1^2 (\ln^2 t - \ln t - 1) + b_0 b_2}{b_0^4 t^2} \right],
\label{Coupling}
\end{align}
with $t= \ln \mu^2/\Lambda^2_{\rm QCD}$ and
\begin{align}
 b_0 &= \frac{33-2n_f}{12\pi},\\
 b_1 &= \frac{153 - 19n_f}{24\pi^2},\\
 b_2 &= \frac{2857 - \dfrac{5033}{9}n_f + \dfrac{325}{27}n_f^2}{128\pi^2},
\end{align}
where $n_f$ represents the number of active flavors.

\begin{table}[b]
\caption{The $\eta_c$-meson leading-twist LCDA moments $\langle \xi^n_{2;{\eta_c}} \rangle$ up to $4_{\rm th}$-order. The errors are squared average of those from all the input parameters, such as the Borel parameter $M^2$ (in unit: GeV$^2$), the condensates and the bound state parameters. The scale $\mu$ is set to be $4~{\rm GeV}$. }
\begin{tabular}{ c c c c c c c c }
\hline
~ & ~$\langle \xi^n_{2;{\eta_c}}\rangle$~ &~$M^2$~   \\
\hline
~$n=2 $~& ~$0.103 \pm 0.009$~ & ~$[2.086,3.078] $~\\
~$n=4 $~& ~$0.031 \pm 0.003$~ & ~$[2.244,3.202]$~\\
\hline
\end{tabular}
\label{Tab:xi}
\end{table}

\section{Numerical results discussions} \label{Section:III}

In doing numerical analysis, we adopt the following parameters. Two nonperturbative gluon condensates are taken as $\langle\alpha_s G^2\rangle = 0.038(11)~{\rm GeV}^4$ and $\langle g_s^3 f G^3\rangle = 0.013(7)~{\rm GeV}^6$~\cite{Colangelo:2000dp, Zhong:2014jla}. The masses of $B_c^+$ and $\eta_c$-meson are taken as $m_{B_c^+} = 6.274~{\rm GeV}$ and $m_{\eta_c} = 2.98 ~ {\rm GeV}$ from Particle Data Group (PDG)~\cite{Zyla:2020zbs}. The $m_{B_c^+}$ has conducted certain research~\cite{Gershtein:1976mv}. The $B_c$-meson decay constant is taken as $f_{B_c} = 489 \pm 5~{\rm MeV}$~\cite{Chiu:2007km}, and the $\eta_c$-meson decay constant is taken as $f_{\eta_c} = 453 \pm 3~{\rm MeV}$~\cite{Zhong:2014fma}.

\subsection{The $\eta_c$ LCDA and the $B_c \to \eta_c$ TFFs}

We first derive the $\eta_c$-meson leading-twist LCDA moments $\langle \xi^n_{2;{\eta_c}} \rangle$ by using the QCD sum rules (\ref{srborel}). Instead of directly computing $\langle \xi^n_{2;{\eta_c}} \rangle$, we derive the $n_{\rm th}$-order moment by using the sum rules for the normalized ratio $f_{\eta_c}^2{\langle \xi^n_{2;{\eta_c}} \rangle}/f_{\eta_c}^2{\langle \xi^0_{2;{\eta_c}} \rangle}$. This approach significantly reduces the theoretical uncertainty, as the sources of uncertainty for $f_{\eta_c}$ and $\langle \xi^n_{2;{\eta_c}} \rangle$ are mutually correlated with each other. Our results are presented in Table~\ref{Tab:xi}, which includes the moments $\langle \xi^n_{2;{\eta_c}} \rangle$ up to $n=4$. Using the usual criteria, {\it e.g.} the continuum contributions are no more than (6\% , 10\%) and the dimension six contribution are less than (0.1\% , 5\%) for $n=(2,4)$, respectively, together with the extra requirement of the flatness of the moments versus $M^2$ ({\it e.g.} as a conservative estimation, we require their uncertainties to be less than $\pm 10\%$ within the allowable windows), we then obtain the Borel windows $M^2 \in [2.086,3.078] \, \text{GeV}^2$ for $\langle \xi^2_{2;{\eta_c}} \rangle$, $M^2 \in [2.244,3.202] \, \text{GeV}^2$ for $\langle \xi^4_{2;{\eta_c}} \rangle$. In the sum rules, the initial scale has been set as the usual choice of $\mu_0=\sqrt{M^2}$, and the moments have been run to $(\mu =\sqrt{M_{B^+_c}^2-m_b^2}\simeq 4{\rm GeV})$ in Table \ref{Tab:xi}. Figure~\ref{Fig:xiM2} demonstrates the stability of the moments within these allowable Borel windows. During the calculation, all sources of uncertainty, including the Borel parameter, the dimension-four condensate $\langle \alpha_s G^2 \rangle$, the dimension-six condensate $\langle g_s^3 f G^3 \rangle$, and the bound state parameters, were considered.

\begin{figure}[t]
\centering
\includegraphics[width=0.42\textwidth]{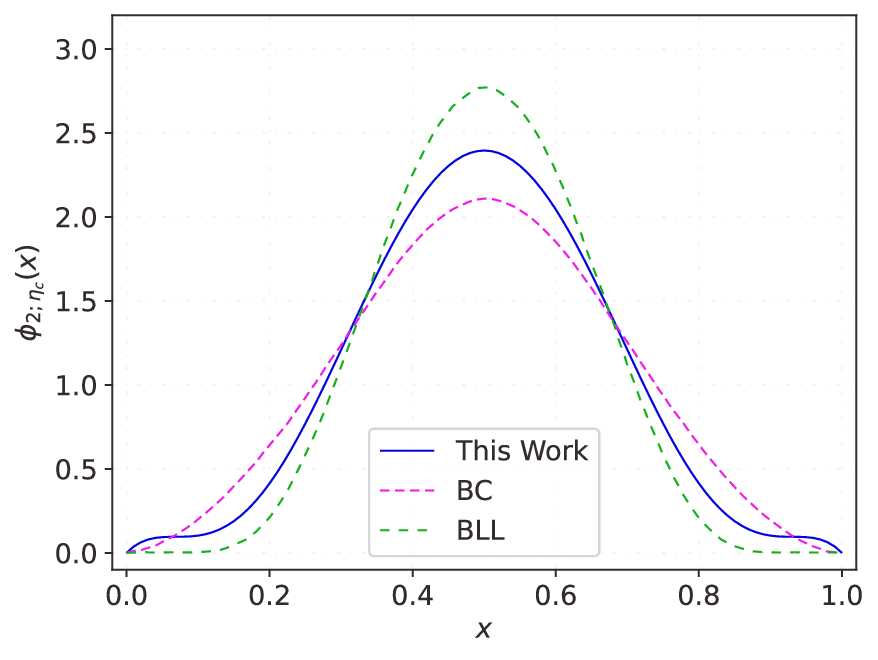}
\caption{The ${\eta _c}$-meson leading-twist LCDA $\phi_{2;\eta _c}(x)$ at the scale $\mu=4~{\rm GeV}$. As a comparison, the BC model~\cite{Bondar:2004sv} and the BLL model~\cite{Braguta:2006wr} have also been presented. } \label{Fig:LCDA}
\end{figure}

Using the relations (\ref{rel_mom2}-\ref{rel_mom4}) among the moments $\langle \xi^n_{2;{\eta_c}} \rangle$ and the Gegenbauer moments $a_n^{\eta_c}$, we obtain $a_2^{\eta_c}=-0.283 \pm 0.026$ and $a_4^{\eta_c}=0.099\pm 0.028$ at the scale $\mu=4$ GeV. Our ${\eta_c}$ meson leading-twist LCDA $\phi_{2;\eta _c}(x)$ at the scale $\mu=4~{\rm GeV}$ is presented in Fig.~\ref{Fig:LCDA}. As a comparison, the results of the BC model~\cite{Bondar:2004sv} and Braguta, Likhoded and Luchinsky (BLL) model~\cite{Braguta:2006wr} have also been presented. Those results close in shape and indicate the $\phi_{2;\eta_c}(x)$ exhibits a symmetric and single-peak behavior.

\begin{table}[htb]
    \centering
    \caption{The $B_c \to \eta_c$ TFFs at ${q^2} = 0$. The errors are squared averages of all mentioned error sources. For comparison, we also present the predictions from various approaches.}
    \label{Tab:TFF0}
    \begin{tabular}{ll}
        \hline
        ~~~~~~~~~~~~~~~~~~~~~~~~~~~~~~~~~~~~~~~~~ & $f_{+}(0)=f_{0}(0)$ \\
        \hline
        This work &  $0.478^{+0.086}_{-0.060}$  \\
        LCSR~\cite{Leljak:2019eyw} &  $0.62^{+0.05}_{-0.05}$  \\
        pQCD~\cite{Wang:2012lrc} & $0.48^{+0.06}_{-0.06}$ \\
        EFG~\cite{Ebert:2003cn} & $0.47$  \\
        HNV~\cite{Hernandez:2006gt} & $0.49^{+0.01}_{-0.01}$  \\
        HQSS~\cite{Biswas:2023bqz} & $0.52^{+0.28}_{-0.28}$ \\
        CLFQM~\cite{Wang:2008xt} & $0.61^{+0.04}_{-0.05}$  \\
        CCQM~\cite{Tran:2018kuv} & $0.75$  \\
        \hline
    \end{tabular}
\end{table}

Regarding the TFFs of $B_c \to \eta_c$, we adopt the following criteria to set the Borel window for the LCSR of these two processes, {\it e.g.} we require the continuum contribution to be less than $30\%$ of the total LCSR and the contribution from high-twist states should be as small as possible. In this work, we adopt the continuum threshold for the TFF as $s_0=42~{\rm GeV}^2$. This value is smaller than one adopted in LCSRs which is calculated by using the usual correlator as reported in Ref.\cite{Chabab:1993nz}. This choice effectively minimizes the contributions from unwanted scalar resonances introduced by the use of the chiral correlator. The Borel windows are determined to be $M^2 = 5.0 \pm 0.5~{\rm GeV}^2$. We present the TFFs at the maximum recoil point $q^2 = 0$ in Table~\ref{Tab:TFF0}, where the errors are the squared averages of all the mentioned error sources for the LCSRs. In Table~\ref{Tab:TFF0}, we also present the results of various groups or approaches, {\it e.g.} Ebert-Faustov-Galkin (EFG) group~\cite{Ebert:2003cn}, Hernandez-Nieves-Verde-Velasco (HNV) group~\cite{Hernandez:2006gt}, heavy quark spin symmetry (HQSS)~\cite{Biswas:2023bqz}, the pQCD approach~\cite{Wang:2012lrc}, the covariant light-front quark model (CLFQM) approach~\cite{Wang:2008xt} and the covariant confined quark model (CCQM) approach~\cite{Tran:2018kuv}.

The LCSR approach for $B_c\to \eta_c$ TFFs are reliable in low and intermediate $q^2$-regions, which can be extrapolated to the physically allowable region $m_\ell^2 \leq q^2 \leq (m_{B_c} - m_{\eta_c})^2$. In this paper, we adopt the simplified series expansion (SSE) to do the extrapolation, {\it e.g.}
\begin{eqnarray}
F_i(q^2) = \frac1{P_i(q^2)}\sum\limits_k \alpha_k [z(q^2)-z(0)]^k,
\end{eqnarray}
where $F_i(q^2)$ represent the two TFFs $f_{0,+}(q^2)$, respectively. Here the function $z(t)$ is defined as
\begin{eqnarray}
z( t) = \frac{{\sqrt {{t_ + } - t}  - \sqrt {{t_ + } - {t_0}} }}{{\sqrt {{t_ + } - t}  + \sqrt {{t_ + } - {t_0}} }},
\end{eqnarray}
where ${t_\pm} = {({m_{B_c}} \pm{m_{\eta_c}})^2}, {t_0} = {t_ + }(1 - \sqrt {1 - {t_ {-}}/{t_{ + }}} )$, whose definition can be found Ref.\cite{Bharucha:2010im}. In this approach, the simple pole $P_i(q^2)=(1-q^2 / m^2_{R,i})$, which accounts for low-lying resonances, is used. The masses of low-lying $B_c$ resonances are mainly determined by their $J^P$ states. Following Refs.\cite{Aad:2014laa, Aaij:2016qlz, Eichten:1994gt, Godfrey:2004ya} and the PDG values~\cite{Zyla:2020zbs}, we list the required $J^P$ states and $m_{R,i}$ in Table~\ref{Tab:SSEfit}. Meanwhile, we introduce the parameter $\Delta$ to measure the quality of fit, which is defined as
\begin{eqnarray}
\Delta  = \frac{\sum\nolimits_t|F_i(t)-F_i^{\rm fit}(t)|}{ \sum\nolimits_t |F_i(t)|}\times  100
\end{eqnarray}
where $t \in [0,1/40, \cdots ,40/40]\times  6.48~{\rm GeV}^2$. The fitting parameters $\alpha_i$ for every TFFs and the quality of fit $\Delta$ are also listed in Table~\ref{Tab:SSEfit}. It shows that the $\Delta$ of $B_c\to {\eta_c}$ TFFs are less than 0.02\%, far less than the usual criterion of $\Delta<1\%$, indicating a high quality of fit. Meanwhile, all these $\Delta$ values for the TFFs' errors are quite small. The small value of $\Delta$ results from both the application of the SSE in the fitting method and the relatively flat variation trend that the TFFs exhibits.

\begin{table}[t]
\centering
\caption{We set all input parameters to their central values and present the masses of low-lying $B_c$ resonances, the coefficients $\alpha_{1,2}$ and $\Delta$ for the TFFs $f_{0,+}(q^2)$.} \label{Tab:SSEfit}
\begin{tabular}{c  c c c c c c }
\hline
TFFs &  $J^P$ & $m_R$  (GeV)  & $\alpha_1$ & $\alpha_2$ & $\Delta$ \\ [0.5ex]
\hline
$f_+^{B_c\to \eta_c}$ & $1^-$ & 6.34  & $-8.12$ & $121.48$ & $0.0183\%$\\
$f_0^{B_c\to \eta_c}$ & $0^+$ & 6.71  & $-2.95$ & $9.59$ & $0.0023\%$\\
\hline
\end{tabular}
\end{table}
\begin{figure}[t]
\centering
\includegraphics[width=0.42\textwidth]{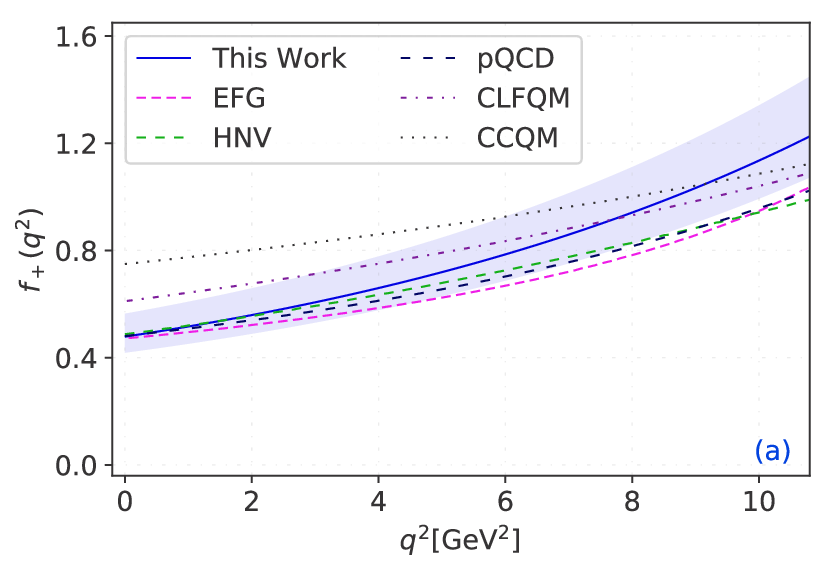}
\includegraphics[width=0.42\textwidth]{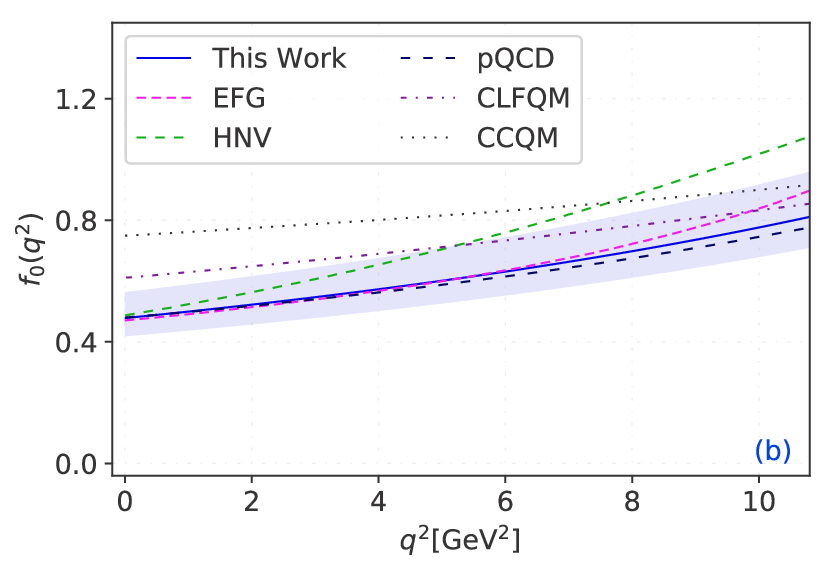}
\caption{Extrapolated LCSR predictions for the $B_c \to \eta_c$ TFFs $f_{+,0}(q^2)$ in whole ${q^2}$-region,in which the shaded bands are squared average of those from the  mentioned error sources. Among them, figure (a) is the $f_{+}(q^2)$, figure (b) is the $f_{0}(q^2)$.} \label{Fig:etacTFFs}
\end{figure}

The predicted $B_c\to \eta_c$ TFFs are shown in Fig.~\ref{Fig:etacTFFs}, where the shaded band shows the error of various input parameters. As a comparison, we also present the results of various groups or approaches in the figure, {\it e.g.} the EFG group~\cite{Ebert:2003cn}, the HNV group~\cite{Hernandez:2006gt}, the pQCD approach~\cite{Wang:2012lrc}, the CLFQM approach~\cite{Wang:2008xt} and the CCQM approach~\cite{Tran:2018kuv}, where only their central values are given. Our results are consistent with most of theoretical calculations within errors.

\begin{figure}[t]
\centering
\includegraphics[width=0.42\textwidth]{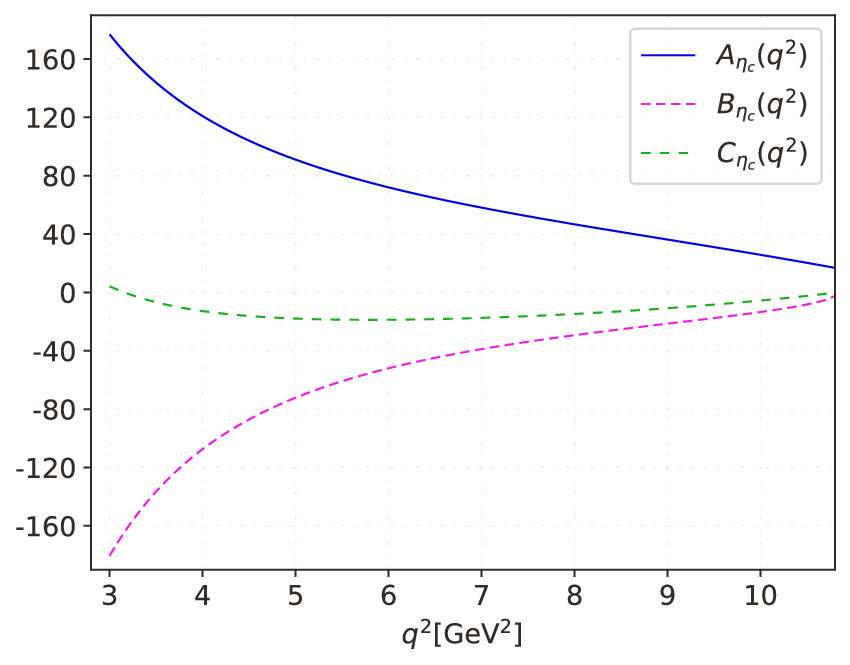}
\caption{The LCSR predictions for the functions $A_{\eta_c}(q^2)$, $B_{\eta_c}(q^2)$ and $C_{\eta_c}(q^2)$ in whole ${q^2}$-region.} \label{ABC:1bot}
\end{figure}

In Fig.~\ref{ABC:1bot}, we present our results of $A_{\eta_c}(q^2)$, $B_{\eta_c}(q^2)$, and $C_{\eta_c}(q^2)$ for the differential distribution (\ref{Eq:dGamma-Bcetac}). It shows that $A_{\eta_c}(q^2)$ provides the dominant contribution in most of $q^2$ region, $B_{\eta_c}(q^2)$ provides a negative contribution, and $C_{\eta_c}(q^2)$ is negligible. Moreover, we observe that $A_{\eta_c}(q^2)$ and $B_{\eta_c}(q^2)$ exhibit significant variations when $q^2$ is small, whereas their changes are relatively minor when $q^2$ is large.

\subsection{The ratio ${R_{\eta_c}}({q^2})$}

Using the TFFs (\ref{eq:fp}) and (\ref{eq:f0}), we obtain the SM prediction for the ratio of $B\to \eta_c$ semileptonic branching fractions,
\begin{eqnarray}
R_{\eta_c}|_{\rm SM}& \equiv\dfrac{\Gamma(B_c\rightarrow \eta_c \tau \bar{\nu}_{\tau})}{\Gamma(B_c\rightarrow \eta_c \mu \bar{\nu}_{\mu})} = 0.308^{+0.084}_{-0.062}.
\end{eqnarray}
Our results are consistent with the results derived under various approaches, which as mentioned in the Introduction vary from $\sim 0.26$ to $\sim 0.38$. At present, there still exists a significant discrepancy between the theoretical predictions and the experimental result regarding the $b \to c$ decays.

\begin{table}[t]
\begin{tabular}{lll}
\hline
~~~~~~~~~~~~~~~~~~~~~~~~~~~~~~~~~~~& $R_{\eta_c}$ ~~~~~~~~~~~~~~~~~~~ & $R_{\eta_c}$~\cite{Leljak:2019eyw}\\
\hline
SM                   & $0.308_{ - 0.062}^{ + 0.084}$  & $0.32$\\
$V_L$                & $0.308_{ - 0.062}^{ + 0.084}$  & $0.39_{ - 0.36}^{ + 0.42}$\\
$S_L$                & $0.365_{ - 0.085}^{ + 0.106}$  & $0.44_{ - 0.33}^{ + 0.55}$\\
$S_R$                & $0.377_{ - 0.088}^{ + 0.111}$  & $0.49_{ - 0.40}^{ + 0.59}$\\
$S_L = 4T_L$         & $0.269_{ - 0.051}^{ + 0.065}$  & $0.26_{ - 0.20}^{ + 0.34}$\\
$(V_L,S_L=  - 4T_L)$ & $0.324_{ - 0.072}^{ + 0.091}$  & $0.42$\\
${({S_{R,}}{S_L})}$  & $0.365_{ - 0.081}^{ + 0.109}$   & $0.45$\\
${({V_{L,}}{S_R})}$  & $0.333_{ - 0.070}^{ + 0.096}$  & $0.44$\\
${\rm Re},{\rm Im} [S_L = 4T_L]$ & $0.275_{ - 0.044}^{ + 0.055}$  & $0.43$\\ \hline
\end{tabular}
\caption{The values of $R_{\eta_c}$ in the presence of SM and various 1D and 2D NP scenarios. The errors are squared average of the ones from all the mentioned input parameters. Results of Ref.\cite{Leljak:2019eyw} are presented as a comparison. }
\label{tab:REtac}
\end{table}

\begin{figure}[t]
\centering
\includegraphics[width=0.42\textwidth]{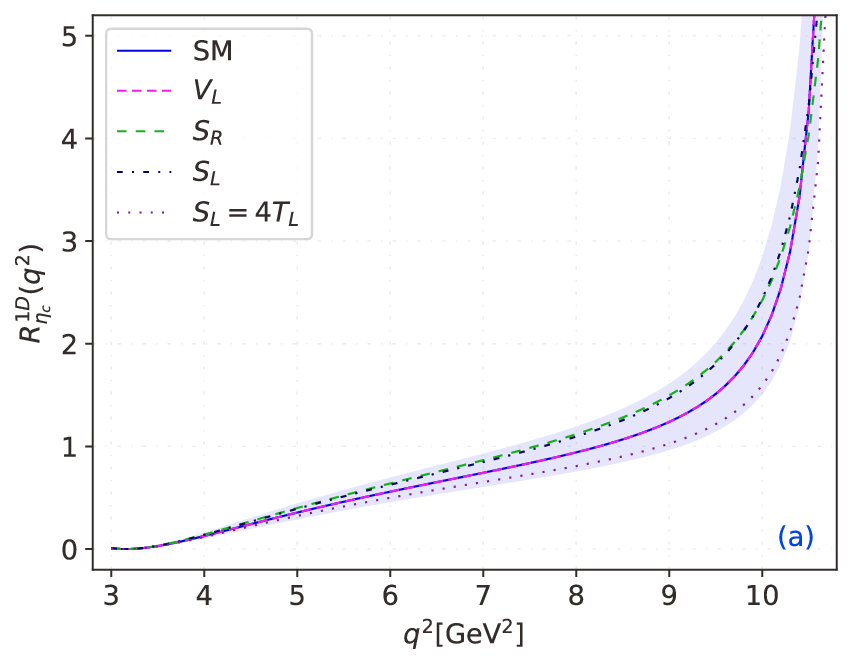}
\includegraphics[width=0.42\textwidth]{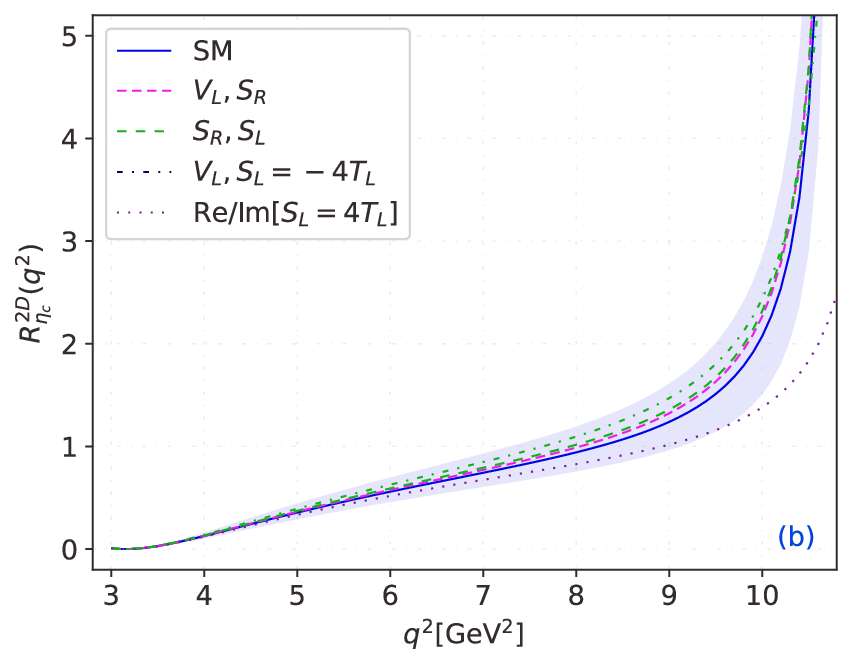}
\caption{The ratio ${R_{\eta_c}}({q^2})$ in the presence of SM and various 1D and 2D NP scenarios using the extrapolated LCSRs. The shaded band is SM error induced by the squared average of the errors of all the input parameters. The blue solid line is the SM prediction, the other lines are SM together with the NP effects for the central parameter values. Among them, figure (a) is the 1D scenario, figure (b) is the 2D scenarios.}
\label{Fig:Ratio}
\end{figure}
Due to sizable decay width, the channel $B_c \to \eta_c\ell\bar\nu_\ell$ may be a helpful platform for searching of NP beyond SM. Regarding the Wilson coefficients listed in Eq.~(\ref{Eq:Lag}), we adopt their possible values given in Ref.\cite{Blanke:2018yud} to do our discussions. The interested readers may turn to it for detailed analyses. For convenience, we present their main results in the following. The possible values of the Wilson coefficients have been determined through a combined analysis of the BABAR, the Belle, and the LHCb data by using two types of scenarios, {\it e.g.} one-dimensional (1D) scenarios and two-dimensional (2D) scenarios. Both types of scenarios are generated by a single new particle added to the SM and the NP coefficients are generated by the exchange of a single heavy spin-0 or spin-1 particle. As for 1D scenarios, only a single Wilson coefficient receives a NP contribution, and the fixed Wilson coefficients at the scale of 1 TeV for four cases are~\cite{Blanke:2018yud},
\begin{eqnarray}
 V_L &=&  0.11\; [0.06, 0.15],  \\
 S_R &=& 0.16\; [0.08, 0.23],  \\
 S_L &=& 0.12\; [0.01, 0.20],   \\
 S_L &=& 4 T_L = -0.07\; [-0.15, 0.02], \label{eq:1Dfit}
\end{eqnarray}
where the central values represent the best fit to the data, and the ranges stand for the fit of data within $2\sigma$ deviation. The coefficients at the scale ({\it e.g.} $\sim m_b$) can be obtained by using renormalization group equation ~\cite{Gonzalez-Alonso:2017iyc}. In their fitting process, only the real values of the coefficients were considered. Furthermore, the potential for allowing imaginary coefficients was explored in Ref.\cite{Angelescu:2018tyl}, which found that the relation ${\rm Im}[S_L] = 4 \,{\rm Im}[T_L]$ is also compatible with the recent experimental data. Consequently, we adopt the best-fit value $S_L = 4 \,T_L$ of Eq.~(\ref{eq:1Dfit}) for both the real and imaginary cases. As for 2D scenarios, two Wilson coefficients are simultaneously affected by NP contributions, and in the following, we only give the central values for the Wilson coefficients at the scale of 1 TeV for four cases, which are~\cite{Blanke:2018yud}
\begin{eqnarray}
&&(V_L, S_L = -4 T_L)  =  (0.08, 0.05),   \\
&&(S_R, S_L) = (-0.30,-0.64),  \\
&&(V_L, S_R) = (0.09, 0.06),  \\
&&({\rm Re}, {\rm Im}[S_L = 4 T_L]) = (-0.06, \pm 0.40), \label{eq:2Dfit}
\end{eqnarray}
which represent the best fit of data.
\begin{table*}[htb]
	\footnotesize
	\begin{tabular}{lllllllll}
		\hline
		& ${\cal A}_{\rm FB}^{\eta_c}$ ~~~~~~~~~~~~~~~& ${\cal C}_F^{\tau,\eta_c}$~~~~~~~~~~~~~~\, & ${\cal P}_L^{\eta_c}$~~~~~~~~~~~~~\, & ${\cal P}_T^{\eta_c}$ ~~~~~~~~~~~~~& ${\cal A}_{\rm FB}^{\eta_c}$\cite{Leljak:2019eyw}~~~~~~~~~ & ${\cal C}_F^{\tau,\eta_c}$\cite{Leljak:2019eyw} ~~~~~~~& ${\cal P}_L^{\eta_c}$\cite{Leljak:2019eyw}~~~~~      & ${\cal P}_T^{\eta_c}$\cite{Leljak:2019eyw}    \\\hline
		SM                         & $-0.370^{+0.013}_{-0.003}$   & $-0.244^{+0.056}_{-0.059}$  & $0.239^{+0.115}_{-0.119}$  & $0.278^{+0.064}_{-0.060}$ & $-0.35$ & $-0.22$ & $0.42$ & $0.81$\\
		$V_L$                      & $-0.370^{+0.013}_{-0.003}$   & $-0.244^{+0.056}_{-0.059}$  & $0.239^{+0.115}_{-0.119}$  & $0.278^{+0.064}_{-0.060}$ & $-$ & $-$ & $-$ & $-$\\
		$S_L$                      & $-0.359^{+0.022}_{-0.012}$   & $-0.202^{+0.051}_{-0.055}$  & $0.313^{+0.106}_{-0.114}$  & $0.350^{+0.061}_{-0.061}$ & $-0.31^{-0.29}_{-0.34}$ & $-0.16^{-0.13}_{-0.21}$ & $0.58^{0.66}_{0.43}$ & $0.73^{0.80}_{0.67}$\\
		$S_R$                      & $-0.356^{+0.023}_{-0.014}$   & $-0.193^{+0.050}_{-0.055}$  & $0.326^{+0.103}_{-0.112}$  & $0.362^{+0.060}_{-0.061}$ & $-0.30^{-0.28}_{-0.32}$ & $-0.14^{-0.12}_{-0.17}$ & $0.62^{0.68}_{0.53}$ & $0.70^{0.76}_{0.66}$\\
		$S_L = 4 T_L$              & $-0.364^{+0.010}_{-0.001}$   & $-0.280^{+0.061}_{-0.062}$  & $0.195^{+0.120}_{-0.122}$  & $0.206^{+0.070}_{-0.062}$ & $-0.36^{-0.34}_{-0.36}$ & $-0.27^{-0.21}_{-0.35}$ & $0.31^{0.45}_{0.14}$ & $0.84^{0.86}_{0.80}$\\
		$(V_L,S_L =  - 4T_L)$      & $-0.360^{+0.018}_{-0.008}$   & $-0.231^{+0.056}_{-0.059}$  & $0.276^{+0.113}_{-0.119}$  & $0.300^{+0.063}_{-0.061}$ & $-0.33$ & $-0.19$ & $0.50$ & $0.77$\\
		$(S_R, S_L)$               & $-0.359^{+0.022}_{-0.012}$   & $-0.202^{+0.051}_{-0.055}$  & $0.313^{+0.106}_{-0.114}$  & $0.350^{+0.061}_{-0.061}$ & $-0.31$ & $-0.15$ & $0.59$ & $0.72$\\
		$(V_L, S_R)$               & $-0.365^{+0.017}_{-0.007}$   & $-0.225^{+0.054}_{-0.058}$  & $0.274^{+0.112}_{-0.117}$  & $0.313^{+0.062}_{-0.060}$ & $-0.33$ & $-0.19$ & $0.50$ & $0.77$\\
		${\rm Re,Im}[S_L = 4T_L]$  & $-0.366^{+0.014}_{-0.003}$   & $-0.234^{+0.055}_{-0.059}$  & $0.229^{+0.110}_{-0.115}$  & $0.265^{+0.065}_{-0.060}$ & $-0.27$ & $-0.16$ & $0.57$ & $0.43$\\\hline
	\end{tabular}
	\caption{The integrated values of the forward-backward asymmetry, the convexity parameter and the longitudinal and transverse polarization of $\tau$, in case of the 1D and 2D NP scenarios discussed in the text. The superscripts and subscripts are induced by the variations of squared average of all input parameters.}
	\label{tab:NP}
\end{table*}

\begin{figure*}[t]
\centering
\includegraphics[width=0.4\textwidth]{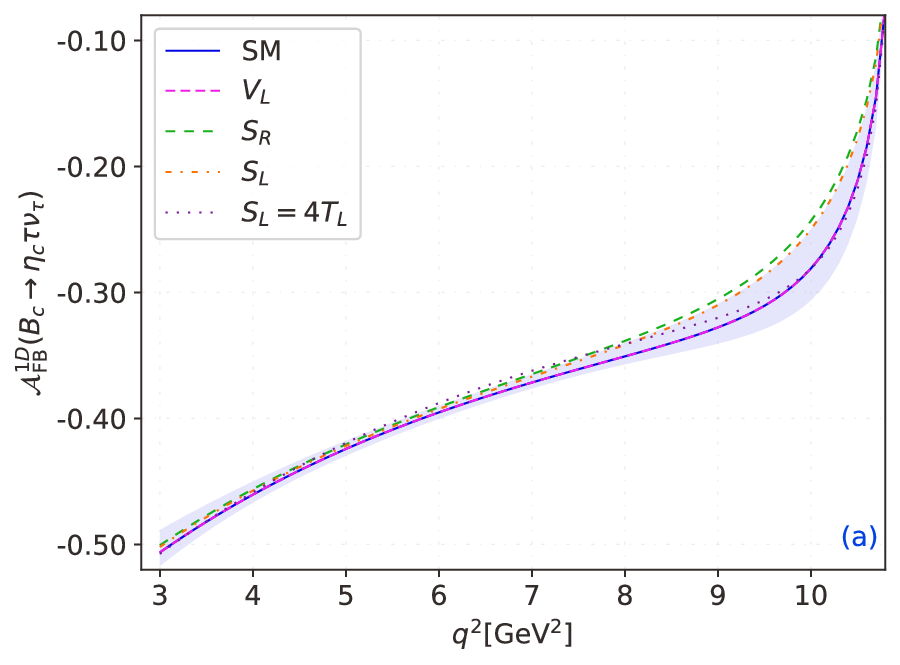}
\includegraphics[width=0.4\textwidth]{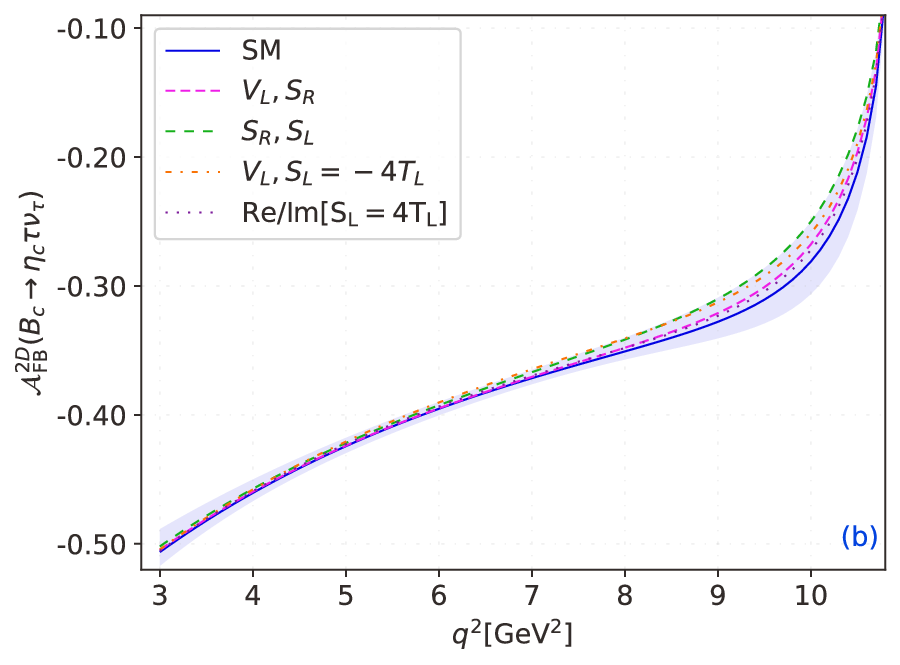}
\includegraphics[width=0.4\textwidth]{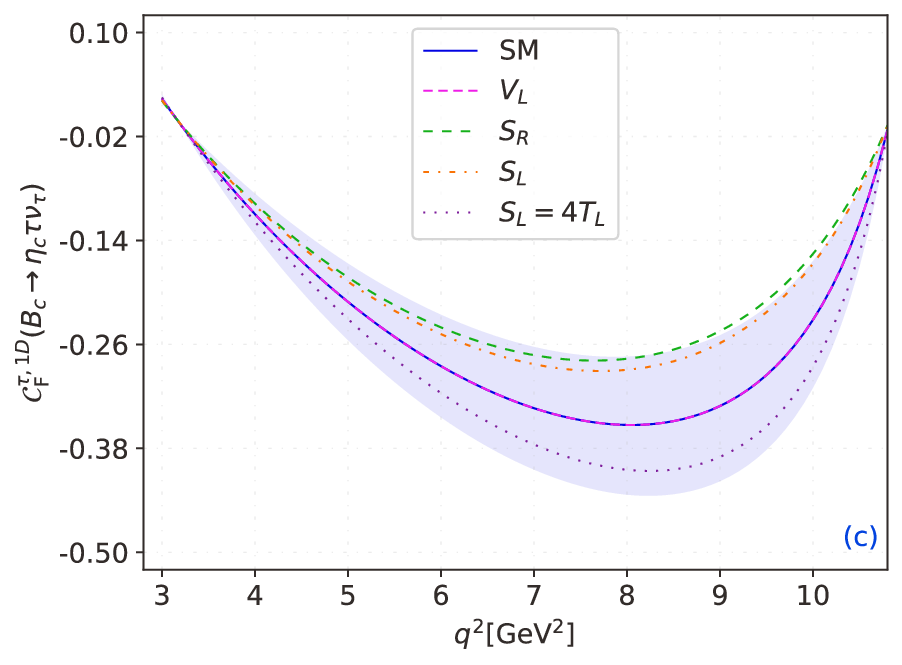}
\includegraphics[width=0.4\textwidth]{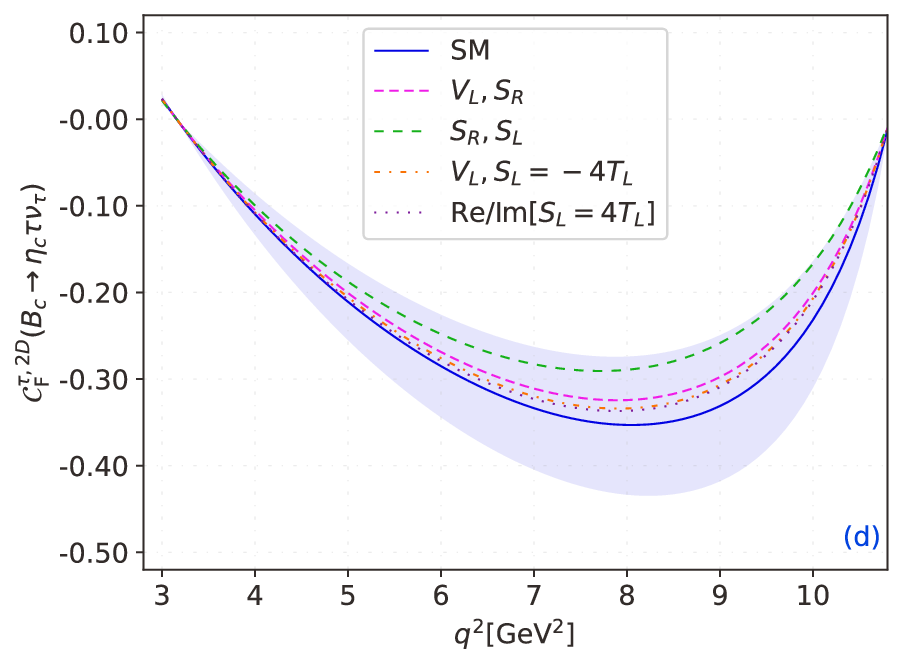}
\caption{The LCSR predictions for the $B_c \to \eta_c$ forward-backward asymmetry and  convexity parameter ${\cal A}_{\rm FB}({q^2})$, ${\cal C}_F^\tau ({q^2})$ in the whole ${q^2}$-region,in which the shaded bands are squared average of those from the  mentioned error sources. The blue solid line is the SM prediction, the other lines are the new physics effects for the best fit points. Among them, figure (a) is the ${\cal A}_{\rm FB}({q^2})$ in the 1D scenario, figure (b) is the ${\cal A}_{\rm FB}({q^2})$ in the 2D scenarios, figure (c) is the ${\cal C}_F^\tau ({q^2})$ in the 1D scenario and figure (d) is the ${\cal C}_F^\tau ({q^2})$ in the 2D scenarios.}
\label{AFB:1bot}
\end{figure*}
\begin{figure*}[t]
\centering
\includegraphics[width=0.4\textwidth]{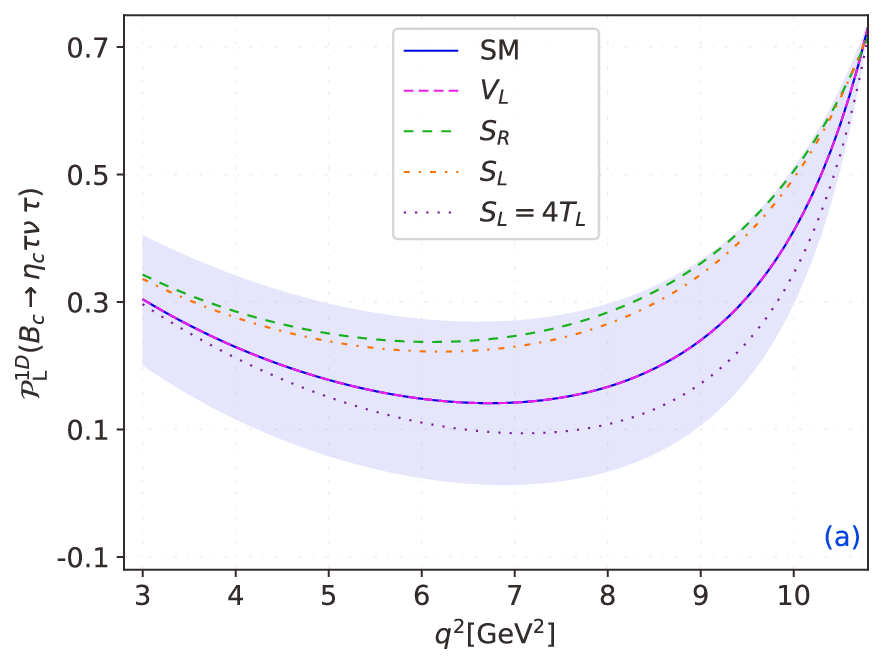}
\includegraphics[width=0.4\textwidth]{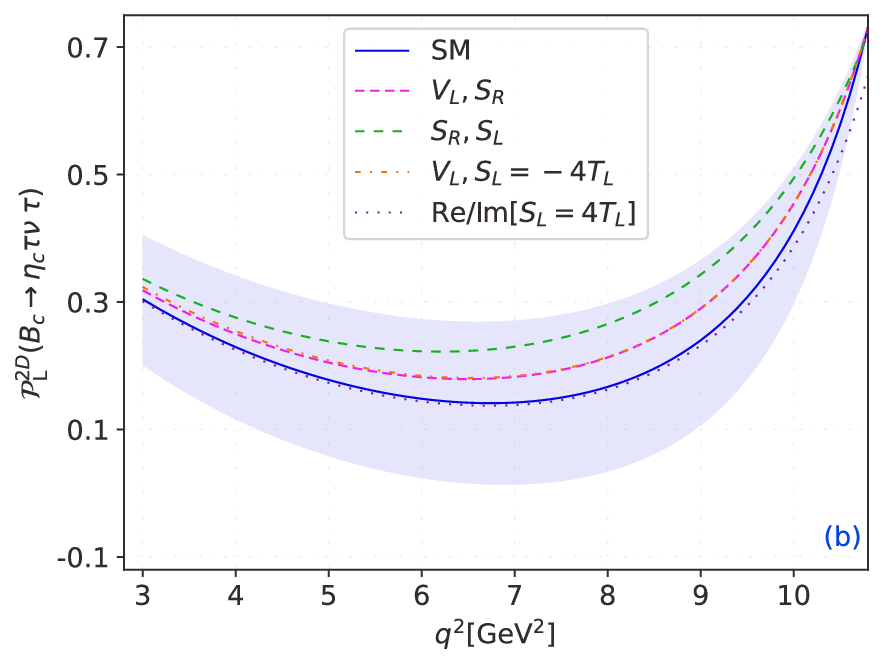}
\includegraphics[width=0.4\textwidth]{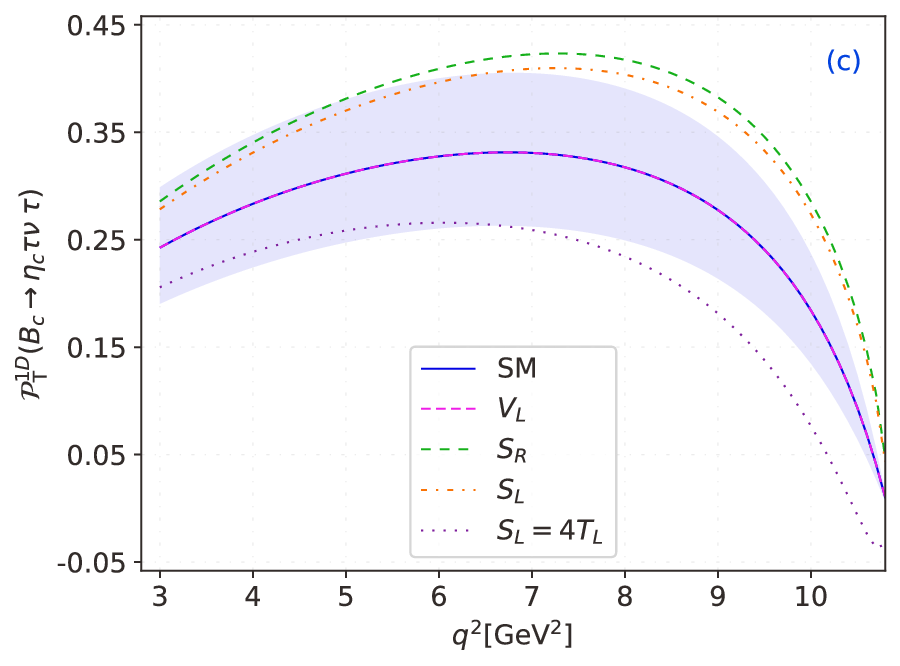}
\includegraphics[width=0.4\textwidth]{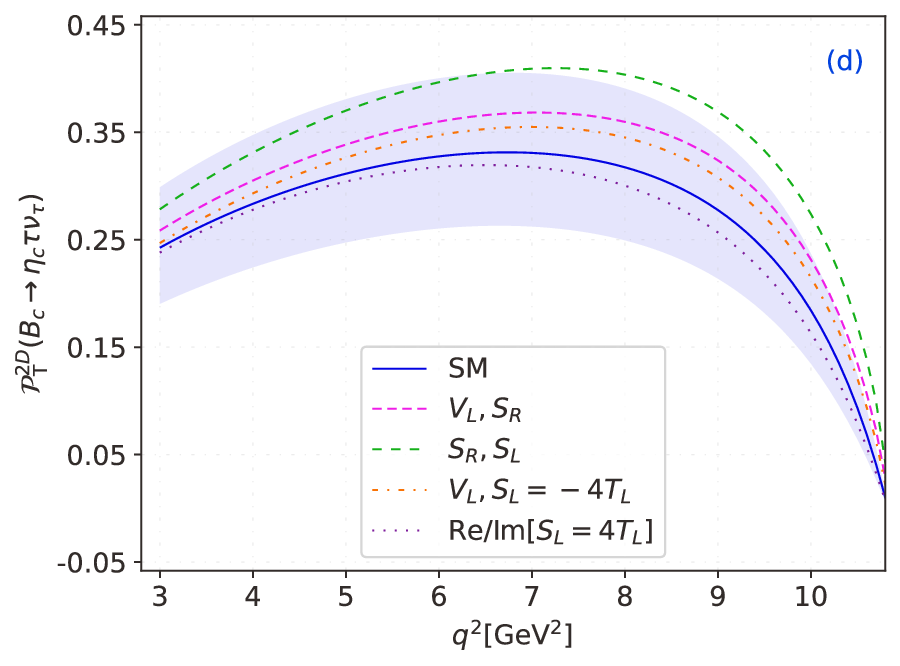}
\caption{The LCSR predictions for the $B_c \to \eta_c$ longitudinal polarization ${{\rm{{\cal P}}}_{L,T}}({q^2})$, in which the shaded bands are squared average of the errors from all the mentioned error sources. The blue solid line is the SM prediction, and the other lines are the NP effects for the best fit points. Among them, figure (a) is the ${{\rm{{\cal P}}}_{L}}({q^2})$ in the 1D scenario, figure (b) is the ${{\rm{{\cal P}}}_{L}}({q^2})$ in the 2D scenarios, figure (c) is the ${{\rm{{\cal P}}}_{T}}({q^2})$ in the 1D scenario and  figure (d) is the ${{\rm{{\cal P}}}_{T}}({q^2})$ in the 2D scenarios.}
\label{PLPT:2bot}
\end{figure*}

Table~\ref{tab:REtac} shows the values of $R_{\eta_c}$ in the presence of SM and various 1D and 2D NP scenarios, respectively, which are integrated results for the entire $q^2$-region. As for the 1D scenarios with nonzero $V_L$ such as the models with vector leptoquarks or left-handed $W_0$ bosons, the ratio will not be changed in comparison to the SM one. This is due to the fact that in this case the nonzero NP factor $|1+V_L|^2$ cancels exactly in the ratio. Furthermore, if the Wilson coefficient has the nonzero terms involving $S_L$, $S_R$, ($V_L,S_L=-4T_L$), ($S_R,S_L$), and ($V_L,S_R$), respectively, the ratio $R_{\eta_c}$ becomes larger. The maximum increase of $R_{\eta_c}$ is $22.4\%$, achieved for the NP models with nonzero $S_R$. However, the $R_{\eta_c}$ decreases when the Wilson coefficient is set to be $S_L=4T_L$ (or $\rm Re,Im[S_L=4T_L]$). The largest decrease of $R_{\eta_c}$ is $12.7\%$, occurring for the case of $S_L=4T_L$.

In Fig.~\ref{Fig:Ratio}, the ratio $R_{\eta_c}$ versus $q^2$ is displayed. For $R_{\eta_c}$ in the 1D scenarios, the NP contributions coincide with the SM prediction in the low and high $q^2$-intervals, with the most significant deviations observed around $q^2=10 ~\rm GeV^2$. Specifically, the scenario where ${S_L} = 4{T_L}$ with purely real or imaginary couplings results in a decrease in the $R_{\eta_c}$ ratios. In the 2D scenarios, the NP contributions generally align with the SM at low $q^2$ values, and most NP contributions lead to an increment of the ratios. Nevertheless, the ($\rm Re,Im[S_L=4T_L]$) scenario with purely real or imaginary couplings results in a decrease in $R_{\eta_c}$. However, the ratios under various NP scenarios are generally within the uncertainty range of those predicted by the SM.

\subsection{The forward-backward asymmetry ${\cal A}_{\rm FB}({q^2})$, convexity parameter ${\cal C}_F^\tau ({q^2})$, longitudinal and transverse polarizations ${{\rm{{\cal P}}}_L}({q^2})$ and ${{\rm{{\cal P}}}_T}({q^2})$}

In Table~\ref{tab:NP}, we show the integrated values of the forward-backward asymmetry, the convexity parameter and the longitudinal and transverse polarization of $\tau$-lepton in the whole $q^2$-region, whose errors are squared average of those from all input parameters. Results of Ref.\cite{Leljak:2019eyw} are presented as a comparison and most of them are consistent with our present predictions within errors.

In Fig.~\ref{AFB:1bot}, we present the forward-backward asymmetry ${\cal A}_{\rm FB}({q^2})$ and convexity parameter ${\cal C}_F^\tau ({q^2})$ of $B_c \to \eta_c$ semileptonic decays. We observe that the forward-backward asymmetry increases with the increment of $q^2$ for all the mentioned NP scenarios. In low to intermediate $q^2$-region, the NP effects are small in comparison to the SM prediction. For the convexity parameter ${\cal C}_F^\tau ({q^2})$, the $S_L = 4 T_L$ case results in a decrease with the increment of $q^2$, while it increases for other cases. However, all NP effects are essentially within the error range of the SM. 

In Fig.~\ref{PLPT:2bot}, we present the longitudinal and transverse polarizations, ${{\rm{{\cal P}}}_L}({q^2})$ and ${{\rm{{\cal P}}}_T}({q^2})$, for $B_c \to \eta_c$ decays. It shows that ${{\rm{{\cal P}}}_L}({q^2})$ will first decrease and then increase with the increment of $q^2$, while for ${{\rm{{\cal P}}}_T}({q^2})$, it will first increase and then decrease with the increment of $q^2$. For the longitudinal and transverse polarizations, the $S_L = 4 T_L$ case results in a decrease, while it increases for other cases.

\section{Summary}\label{Section:IV}

Experimental measurements of semileptonic decays of the $B$ mesons have led to intriguing experimental tensions with the SM in recent years. The LHCb measurement of $B_c \to J/\psi l \nu_l$ decays has led to the speculation on whether the observed potential lepton flavor universality violation in $B$ decays can also be seen in the $B_c$-meson semileptonic decay channels.

In the present paper, we have made a detailed discussion on the $B_c \to \eta_c$ semileptonic decay. Its dominant components, {\it e.g.} the $B_c \to \eta_c$ TFFs, have been calculated by using the LCSR approach and then extrapolated to full $q^2$ region via SSE. The NLO QCD corrections to the TFFs have also been discussed. And to improve the precision of LCSRs, we have also recalculated the $\eta_c$-meson leading-twist LCDA by using the BFTSR approach. Our results of TFFs are in agreement with previous predictions within reasonable errors. The SM branching ratios of the $B_c$-meson to $\eta_c$ has been calculated and compared with the results from other approaches. Moreover, our SM prediction of the semileptonic ratio is $R_{\eta_c}|_{\rm SM} = 0.308^{+0.084}_{-0.062}$.

Till now, none of the NP scenarios have been confirmed via the global fit of available experimental data on the $B \to (D,D^{\ast}) \ell \nu_\ell$ semileptonic decays~\cite{Blanke:2018yud} or from the current 2$\sigma$ tension for $R_{J/\psi}$. We have also discussed the possibility of NP effects in $B_c \to \eta_c$ semileptonic decay based on the effective Hamiltonian approach consisting of the possible four Fermi operators. The constraints on those NP operators have been discussed in Ref.\cite{Blanke:2018yud} from a combined fit of the data on $R_{D(\ast)}$, the $\tau$ and $D^\ast$ longitudinal polarization from $B \to D^{\ast}$ decay and the leptonic $B_c \to \tau \mu$ branching ratios. By using their constraints on the NP parameters under 1D scenarios and 2D scenarios, we analyze the effects of the NP operators on various observables, as shown by Tables (\ref{tab:REtac}, \ref{tab:NP}) and Figs. \ref{Fig:Ratio}, \ref{AFB:1bot}, \ref{PLPT:2bot}. Our results show that most of the NP effects are within the error bands of the SM predictions. On the one hand, we need to further improve theoretical predictions, and on the other hand, we need more precise data such as those from the future HL-LHC on the mentioned observables so as to confirm whether there is NP signal in $B_c \to \eta_c$ semileptonic decays. More data are also helpful to confirm whether there is lepton flavor violation effects in $B_c$-meson decays.

\acknowledgments

H-B. F. and T. Z. would like to thank the Institute of High Energy Physics of Chinese Academy of Sciences for their warm and kind hospitality. This work was supported in part by the National Natural Science Foundation of China under Grants No.~12265010, No.~12265009, No.~12175025, and No.~12347101, the Project of Guizhou Provincial Department of Science and Technology under Grants No.MS[2025]219, No.CXTD[2025]030, No.ZK[2023]024, the Graduate Research and Innovation Foundation of Chongqing, China under Grant No.CYB23011 and No.ydstd1912, the Fundamental Research Funds for the Central Universities under Grant No.~2020CQJQY-Z003.

\end{document}